\documentclass[acmsmall]{acmart}
\AtBeginDocument{%
  }

\setcopyright{cc}
\setcctype{by}
\acmDOI{10.1145/3729405}
\acmYear{2025}
\acmJournal{PACMSE}
\acmVolume{2}
\acmNumber{FSE}
\acmArticle{FSE135}
\acmMonth{7}
\received{2024-09-13}
\received[accepted]{2025-04-01}

\usepackage{tabularx} 
\usepackage{amsmath,amsfonts}
\usepackage{algorithmic}
\usepackage{graphicx}
\usepackage{textcomp}
\usepackage{multirow}
\usepackage{booktabs}
\usepackage{enumitem}
\usepackage{soul}
\usepackage{ulem}
\usepackage{diagbox} 
\usepackage[most]{tcolorbox}  
\usepackage{color}
\usepackage{xcolor}
\usepackage{tikz}
\usepackage{tikz-qtree}
\usepackage{pgfplots}
\usepackage{varwidth}
\usepackage{flushend}
\usetikzlibrary{patterns}
\usepackage{url}
\usepackage{subfigure}
\usepackage{threeparttable}
\usepackage{colortbl}
\usepackage{balance}
\usepackage{bbding}
\usetikzlibrary{plotmarks}
\usetikzlibrary{patterns}
\usepackage{pgfplots}

\begin{document}

\title{Smaller but Better: Self-Paced Knowledge Distillation for Lightweight yet Effective LCMs}




\author{Yujia Chen}
\email{yujiachen@stu.hit.edu.cn}
\affiliation{%
  \institution{Harbin Institute of Technology}
  \city{Shenzhen}
  \country{China}
}

\author{Yang Ye}
\email{yeyang14@huawei.com}
\affiliation{%
  \institution{Huawei Cloud Computing Technologies Co., Ltd.}
  \city{Shenzhen}
  \country{China}
}

\author{Zhongqi Li}
\email{lizhongqi7@huawei.com}
\affiliation{%
  \institution{Huawei Cloud Computing Technologies Co., Ltd.}
  \city{Shenzhen}
  \country{China}
}

\author{Yuchi Ma}
\email{mayuchi1@huawei.com}
\affiliation{%
  \institution{Huawei Cloud Computing Technologies Co., Ltd.}
  \city{Shenzhen}
  \country{China}
}

\author{Cuiyun Gao}
\authornote{Corresponding author.}
\email{gaocuiyun@hit.edu.cn}
\affiliation{
  \institution{Harbin Institute of Technology}
  \city{Shenzhen}
  \country{China}
}



\pgfplotsset{compat=1.18,
             /pgfplots/ybar legend/.style={
              /pgfplots/legend image code/.code={
              \draw [#1,yshift=-0.4em] rectangle (0.7cm,0.2cm);
              },
             }
}

\newcommand{\tool}{SODA}
\newcommand{\model}{SodaCoder}
\newcommand{\firststage}{Correct-and-Fault Knowledge Delivery}
\newcommand{\secondstage}{Multi-View Feedback}
\newcommand{\thirdstage}{Feedback-based Knowledge Update}

\newcommand\etal{{\it{et al.\ }}}
\definecolor{light-gray}{gray}{0.9}    
\definecolor{mygray}{gray}{0.8}    
\definecolor{deep-gray}{gray}{0.7}    
\newcommand{\lgg}{\cellcolor{light-gray}}
\newcommand{\g}{\cellcolor{mygray}}
\newcommand{\hgg}{\cellcolor{deep-gray}}

\definecolor{faulty}{HTML}{AEDFAC}
\definecolor{score}{HTML}{F8C701}
\definecolor{execution}{HTML}{DE757B}
\definecolor{classify}{HTML}{8AAAD6}
\definecolor{base}{HTML}{DADADA}

\definecolor{top5}{gray}{0.9}    
\definecolor{top4}{gray}{0.8}    
\definecolor{top3}{gray}{0.9}    
\definecolor{top2}{gray}{0.8}    
\definecolor{top1}{gray}{0.7}    

\newcommand{\modify}[1]{\textcolor{black}{#1}}

\newcommand{\add}[1]{\textcolor{black}{#1}}

\newcommand{\zhou}[1]{\textcolor{magenta}{[zhou: #1]}}
\begin{abstract}
Large code models (LCMs) have remarkably advanced the field of code generation.
Despite their impressive capabilities, they still face practical deployment issues, 
such as high inference costs,
limited accessibility of proprietary LCMs, and adaptability issues of ultra-large LCMs. These issues 
highlight the critical need for more accessible, lightweight yet effective LCMs. Knowledge distillation (KD) offers a promising solution, which transfers the programming capabilities of larger, advanced LCMs (\textit{Teacher}) to smaller, less powerful LCMs (\textit{Student}). However, existing KD methods for code intelligence often lack consideration of fault domain knowledge and rely on static seed knowledge, leading to degraded programming capabilities
of student models.    

%
In this paper, we propose a novel \textbf{S}elf-Paced 
kn\textbf{O}wledge \textbf{D}istill\textbf{A}tion framework, named {\tool}, aiming
at developing lightweight yet effective student LCMs via adaptively transferring the programming capabilities from advanced teacher LCMs.
{\tool} consists of three stages in one cycle: (1) \textbf{\firststage} stage aims at improving the student model's capability to recognize errors while ensuring its basic programming skill 
during the knowledge transferring, which involves correctness-aware supervised learning and fault-aware contrastive learning methods.
(2) \textbf{\secondstage} stage aims at measuring the quality of results generated by the student model from two views, including model-based and static tool-based measurement, for identifying the difficult questions;
(3) \textbf{\thirdstage} stage aims at updating the student model adaptively by generating new questions at different difficulty levels, in which the difficulty levels are categorized based on the feedback in the second stage.
By performing the distillation process
iteratively, the student model 
is continuously refined through
learning more advanced programming skills from the teacher model.
We compare {\tool} with four state-of-the-art KD approaches on three widely-used code generation benchmarks with different programming languages. Experimental results show that {\tool} improves the student model by 65.96\% in terms of average Pass@1, outperforming the best baseline PERsD by 29.85\%. Based on the proposed {\tool} framework, we develop {\model}, a series of lightweight yet effective LCMs with $\sim$7B parameters, which outperform 15 LCMs with less than or equal to 16B parameters. Notably, {\model}-DS-6.7B, built on DeepseekCoder-6.7B, even surpasses the prominent ChatGPT on average Pass@1 across seven programming languages (66.4 vs. 61.3).

\end{abstract}




\maketitle

\section{Introduction} \label{sec:intro}
    \begin{figure}
    \centering
    \subfigure[The typical process of knowledge distillation.]{
\includegraphics[width=0.6\linewidth]{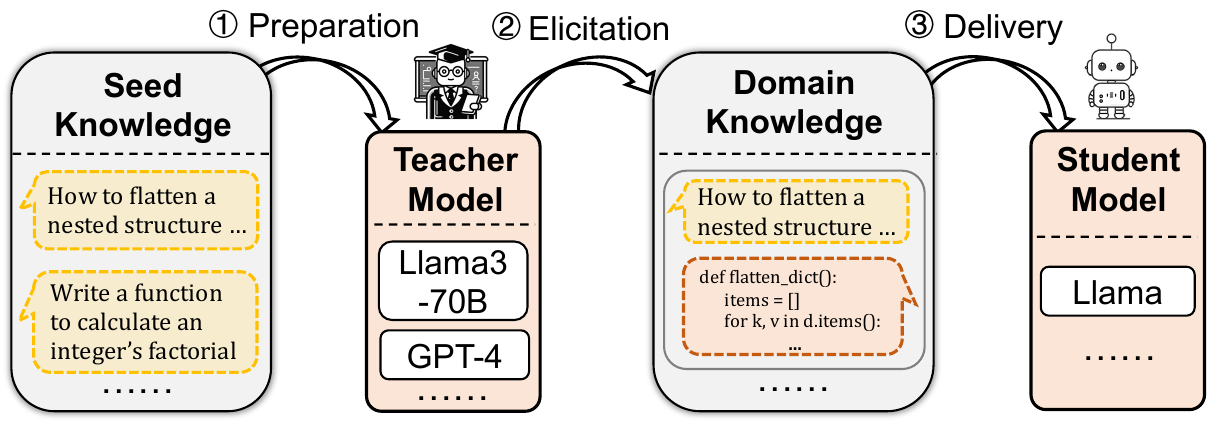}
      \label{fig:motivation_1}
   }
    \subfigure[An example from CodeLlama-7B, distilled by OSS-Instruct.]{
\includegraphics[scale=0.42]{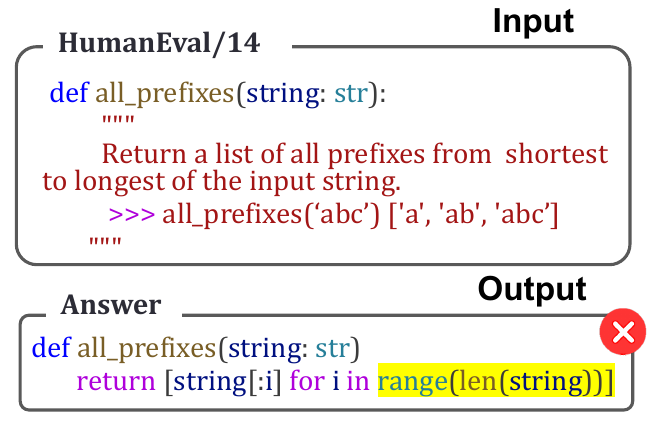}
      \label{fig:motivation_2}
   }
    \caption{Illustration of the typical process of KD
    and a motivating example.}
    \label{fig:motivation}
\end{figure}

In recent years, large language models (LLMs), such as GPT-4~\cite{GPT-4} and ChatGPT~\cite{Chatgpt} have advanced natural language processing (NLP) field; large code models (LCMs), such as CodeLlama~\cite{CodeLlama} and DeepSeek-Coder~\cite{deepseek-coder}, have shown remarkable success in code intelligence field~\cite{code-intelligence-1,code-intelligence-2,code-intelligence-3,code-intelligence-4}, demonstrating their great potential in understanding and generating source code. Despite their impressive capabilities, utilizing these models in practical environments still poses
several challenges. 
Proprietary LLMs like GPT-4 offer accessible APIs, but their high costs and limited accessibility restrict their use among individuals and smaller organizations~\cite{gpt-limits}. In contrast, open-source LCMs like CodeLlama are more accessible and versatile; however, they also have drawbacks. Ultra-large models such as CodeLlama-70B require extensive computational resources to deploy locally~\cite{depoy-locally}, whereas smaller versions such as CodeLlama-7B, though less resource-intensive, often underperform in downstream 
tasks~\cite{underperform-1,underperform-2}. These challenges highlight the pressing need for more accessible, 
effective, yet lightweight LCMs. 
Knowledge distillation (KD), which uses the advanced LLMs (\textit{Teacher}) to enhance the less powerful LLMs (\textit{Student}), offers a promising solution~\cite{KD-survey,minillm,self-instruct,WizardCoder}. 
As shown in Figure~\ref{fig:motivation_1}, a typical process of KD has three phases: (1)
Seed knowledge preparation. This seed knowledge usually comprises specific questions 
to
elicit domain knowledge from the teacher model. (2) Domain knowledge elicitation. In response to the questions in the seed knowledge, the teacher model generates corresponding responses, 
which are expected to encapsulate the teacher's powerful domain skills. (3)
Knowledge delivery. With the elicited domain knowledge, the student model acquires similar capabilities by minimizing the discrepancy between its outputs and those of the teacher model.
In the field of code intelligence, KD methods~\cite{self-instruct, WizardCoder, personalized, Magicoder, mftcoder} focus on transferring programming skills from a teacher model to a student model. 
For example, SELF-INSTRUCT~\cite{self-instruct} uses ChatGPT to create 20,000 code instruction-response pairs from 21 seed tasks, and then employs this synthetic data to fine-tune the student models.   
%
%
OSS-INSTRUCT~\cite{Magicoder} tries to incorporate open-source code into the data generation process, which can guide the teacher model to create more realistic data.
%
However, these KD approaches present two main
limitations.

\textit{1) Lack of consideration of fault domain knowledge:}
In the human learning process, students often learn from mistakes to avoid falling into the same trap twice. However, existing KD approaches focus on improving programming skills by delivering only correct knowledge to the student model. Neglecting the benefits of learning from errors, 
these approaches 
hinder the student's ability to effectively identify and rectify errors in its own solutions, which are crucial for solving programming problems.
As shown in Figure~\ref{fig:motivation_2}, the student model incorrectly employs ``\texttt{range(len(string))}'', not realizing that it excludes the final character index. 
This example indicates
a critical gap: The student model knows to use the API ``\texttt{range()}'' 
but does not fully grasp its parameter's usage ( i.e., ``\texttt{range(1,len(string)+1)}'')
without hints for
common error patterns.

\textit{2) Utilizing static seed knowledge}: 
All existing KD methods
rely on fixed seed knowledge during the entire 
distillation 
process, and cannot update dynamically based on the student model’s current programming skill, which potentially leads to over-training in some specific
areas and under-training in others. 
%
Without updating the seed knowledge, the student model's performance would be limited. 
%
%

\noindent \textbf{Our work.} To mitigate these limitations, we propose {\tool}, a \textbf{S}elf-paced kn\textbf{O}wledge \textbf{D}istill\textbf{A}tion framework. {\tool} consists of three stages in one distillation cycle: 
1) \textit{\firststage} stage. In this stage, the student model learns how to correctly solve programming questions from the teacher model.
Specifically, this stage includes two modules, correctness-aware supervised learning and fault-aware contrastive learning. The former aims to establish the student's basic programming skills by delivering the correct knowledge. The latter helps the student model to identify common programming mistakes by delivering both correct and faulty knowledge, thus further improving its ability to provide accurate solutions.
2) \textit{\secondstage} stage. 
This stage evaluates the quality of solutions generated by the student model from two views, including model-based scoring and static tool-based execution, for identifying the difficult questions. Specifically, in {\it model-based scoring}, we
fine-tune CodeLlama-7B with a small-scale annotated dataset to 
evaluate the solutions' qualities, focusing on widely-adopted
code review standards such as basic functionality, boundary check, readability, etc.
In {\it static tool-based execution}, the solutions are executed in a workspace that supports six different programming languages, and the execution results are categorized as ``Pass'' or ``Fail''.
3) \textit{\thirdstage} stage. 
This stage generates new seed knowledge
for the next distillation cycle by first categorizing the current questions into three difficulty levels (easy, medium, and hard) according to
the feedback in the second
stage, and then prompting the teacher model to set new questions based on the categorized questions. 
By performing the distillation cycle iteratively, the student model is continuously refined through distinguishing errors and learning more advanced programming skills from the teacher model. 

We compare {\tool} with four state-of-the-art KD approaches, including EVOL-INSTRUCT~\cite{WizardCoder}, OSS-INSTRUCT~\cite{Magicoder}, PERsD~\cite{personalized} and MFT~\cite{mftcoder} on three widely-used code generation benchmarks with different programming languages, including HumanEval~\cite{humaneval} and MBPP~\cite{mbpp} for Python, and MultiPL-E~\cite{MultiPL-E} for Java, JavaScript (JS), C, C++, Go and TypeScript (TS).
Experimental results show that {\tool} improves the student model by 84.18\% on HumanEval, 39.52\% on MBPP, 69.27\% on MultiPL-E in terms of the average Pass@1, outperforming the best baseline approach PERsD by 39.90\% and 35.03\% and 30.60\%, respectively.

Based on the proposed {\tool} framework, we develop {\model}, a series of lightweight yet effective LCMs, which contains {\model}-CL and {\model}-DS built upon CodeLlama-7B and DeepseekCoder-6.7B, respectively. Compared with 6 twice-sized LCMs and 9 comparable-sized LCMs, {\model} achieves the highest Pass@1 score at 74.3 on HumanEval. Moreover, {\model}-DS achieves an average Pass@1 at 66.4 across the seven programming languages, even surpassing the prominent ChatGPT (61.3).

\noindent \textbf{Contributions.}
In summary, our main contributions to this paper are as follows:

\begin{itemize}

\item We introduce {\tool}, a self-paced knowledge distillation framework that adaptively 
transfers the programming capability from larger, more advanced LLMs to smaller, less powerful LLMs. Based on the proposed framework, we develop {\model}, a series of lightweight yet effective LCMs with $\sim$ 7B parameters.

\item We propose a novel knowledge delivery approach combining correctness-aware supervised learning and fault-aware contrastive learning to improve the student model's programming skills; and an adaptive knowledge update method
based on multi-view feedback to continuously refine the student model. 

\item We conduct extensive experiments across different benchmarks
in the code generation task. Experimental results show that {\tool} substantially outperforms existing KD approaches and {\model} surpasses 15 state-of-the-art LCMs with less than or equal to 16B parameters. 
Notably, {\model}-DS, built on DeepseekCoder-6.7, achieves an impressive average pass@1 score at 66.4, even surpassing the prominent ChatGPT (61.3).


\end{itemize}
    
\section{Background \& Related Work} \label{sec:background}

\subsection{Large Code Models}

The use of large language models (LLMs) in natural language processing has inspired 
researchers and companies to develop large code models (LCMs) for code-related tasks, especially for code generation. OpenAI proposes Codex~\cite{CodeX}, which is the earlier representative work of LCMs. Meta proposes InCoder~\cite{incoder} and CodeLlama~\cite{CodeLlama}, Salesforce proposes CodeGen~\cite{codegen2}, BigCode project proposes StarCoder series~\cite{starcoder,StarCoder2}, and Deepseek AI proposes DeepseekCoder~\cite{deepseek-coder}. Generally, the pre-trained LCMs can be obtained through two types of approaches: continuously training foundation LLMs or directly training from scratch~\cite{CodeGeeX,CodeGen,CodeLlama,starcoder,deepseek-coder,CodeX}. CodeLlama~\cite{CodeLlama} is an example of the former type, which is trained on LLaMA 2~\cite{llama2} with 0.5 trillion code-specific tokens. DeepseekCoder~\cite{deepseek-coder} is an example of the latter type, which is trained on two trillion tokens across over 80 programming languages. To improve the generalization capacities of these pre-trained LCMs, researchers further fine-tune them on high-quality instruction datasets, as showcased by WizardCoder~\cite{WizardCoder}, MFTCoder~\cite{mftcoder}, Magicoder~\cite{Magicoder}, and WaveCoder~\cite{wavecoder}. The emergence of these models yields remarkable enhancements in the
effectiveness of
code generation, where they automatically generate code snippets in response to given programming questions. To evaluate the performance of LCMs in the code generation task, various benchmarks are developed. Notable examples of these benchmarks include HumanEval~\cite{humaneval}, MBPP~\cite{mbpp}, and MultiPL-E~\cite{MultiPL-E}. These benchmarks typically include programming questions in various languages, accompanied by test cases, and the performance of LCMs is generally measured by the pass rate (Pass@k) of these test cases. In this paper, we develop a series of smaller but better LCMs, named {\model}, which shows superior performance compared to existing LCMs, such as CodeLlama, DeepseekCoder, WizardCoder and MFTCoder, and  even surpasses
the prominent ChatGPT.

\subsection{Knowledge Distillation for LLMs}~\label{sec:back_kd}

In the area of LLMs, knowledge distillation (KD) aims to extract and transfer the powerful capabilities of LLMs (such as ChatGPT) into some smaller and less powerful models (such as Llama-7B). According to the work~\cite{kd-survey-new}, current LLM-based KD methods can be broadly categorized into three types: in-context learning (ICL), chain-of-thought (CoT), and instruction following (IF). 

\textit{ICL distillation}~\cite{icl1,icl2,icl3} focuses on enhancing the student model's few-shot learning abilities by transferring context-specific knowledge from the teacher model. This method leverages the teacher’s ability to perform effectively in various scenarios with minimal training examples. For example, LLM-R~\cite{LLM-R} utilizes an LLM to retrieve high-quality contextual examples, which are then ranked to generate training data for knowledge distillation. \textit{CoT distillation}~\cite{cot1,cot2,cot3,cot4,cot5} leverages the teacher model to generate high-quality rationales for complex reasoning tasks, such as math problems, thereby improving the student model's ability to analyze step-by-step processes. For example, Chen~\textit{et al.}~\cite{MCC-KD} introduce Multi-CoT Consistent Knowledge Distillation (MCC-KD), which uses LLMs to generate multiple fundamental principles for each question and minimize the consistency between corresponding outputs. On the other hand, SCOTT~\cite{SCOTT} introduces the core principle of leveraging an LLM to guide the correct answer through comparative decoding. 
\textit{IF distillation}~\cite{self-instruct,wavecoder,WizardCoder,mftcoder,personalized,Magicoder} enhances the student model's ability to generalize across unseen tasks by training it on diverse task-specific instructions and responses generated by the teacher model. For example, SELF-INSTRUCT~\cite{self-instruct} and then employs 20,000 code instruction-response pairs to fine-tune the student models. EVOL-INSTRUCT~\cite{WizardCoder} creates diverse instruction data from GPT-3.5 for code generation via evolution heuristics. PERsD~\cite{personalized} introduces a personalized distillation approach by asking the teacher model to provide an adaptive refinement for the student model. OSS-INSTRUCT~\cite{Magicoder} guides the teacher model to create more realistic data by incorporating open-source code into the data generation process. MFT~\cite{mftcoder} adapts the student model to multiple code-related tasks through a multi-task fine-tuning framework. 

In the field of code intelligence, IF distillation methods are predominantly used since they align closely with the requirements of coding tasks that often involve understanding and implementing specific requirements, such as generating code snippets. Unlike ICL distillation which improves responses based on context or CoT distillation which builds reasoning path, IF distillation fine-tunes models to respond accurately to explicit programming questions, which have been proven to be highly effective at boosting the capabilities of LCMs~\cite{Magicoder,WizardCoder,mftcoder,personalized}. Therefore, in this paper, we focus on comparing proposed {\tool} with the state-of-the-art IF distillation methods, including EVOL-INSTRUCT~\cite{WizardCoder}, PERsD~\cite{personalized}, OSS-INSTRUCT~\cite{Magicoder} and MFT~\cite{mftcoder}. Different from these methods, we incorporate fault domain knowledge delivery and adaptive seed knowledge update for continuously refining the student model.

\section{Proposed Framework} \label{sec:approach}







\begin{figure*}[t]
\setlength{\abovecaptionskip}{-0.01cm}
    \centering
    \includegraphics[scale=0.38]{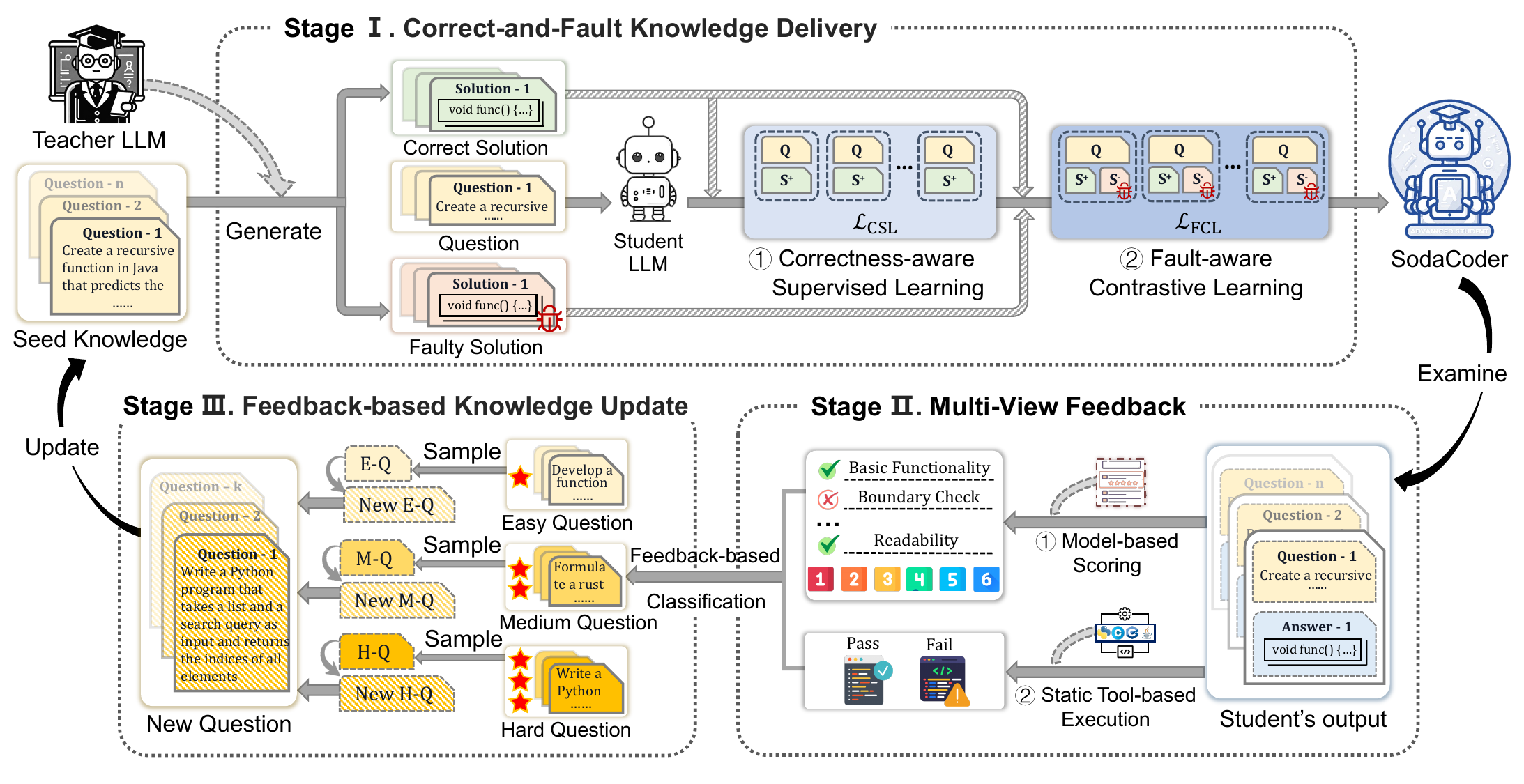}
    \caption{The overview of {\tool}.}
    \label{fig:framework}
\end{figure*}

In this section, we introduce the proposed {\tool}, a self-paced knowledge distillation framework that aims at adaptively transferring
the programming capability from larger,
advanced LLMs (\textit{Teacher}) to smaller, less powerful LLMs (\textit{Student}). We first present the overview of {\tool} and then describe the
details in the following subsections.

\subsection{Overview}

In {\tool}, the teacher and 
student model
interact with each other in a cyclical process
to enable effective knowledge transfer. Each cycle consists of three main stages, as shown in Figure~\ref{fig:framework}.

\begin{itemize}
    \item[(I)] \textit{{\firststage}}: This stage aims to establish the student model's basic programming skills through correctness-aware supervised learning and improve its error identification skills through fault-aware contrastive learning. 
    
    \item[(II)] \textit{{\secondstage}}: This stage evaluates the outputs of the student model from two perspectives: model-based scoring and static tool-based execution, aiming at identifying the difficult questions
    
    
    
    \item[(III)] \textit{{\thirdstage}}: This stage aims to adaptively update the seed knowledge for the subsequent distillation
    cycle based on the feedback in the second
    stage.
    
    
\end{itemize}

\subsection{\firststage}

As shown in Stage I of Fig.~\ref{fig:framework}, the teacher model first generates both correct and faulty solutions for the programming problems in the seed knowledge. To develop basic programming skills for the student model while improving its error identification skills, the correctness-aware supervised learning method and fault-aware contrastive learning method are successively employed.

\begin{figure}[t]
\setlength{\abovecaptionskip}{-0.01cm}
    \centering
    \includegraphics[scale=0.38]{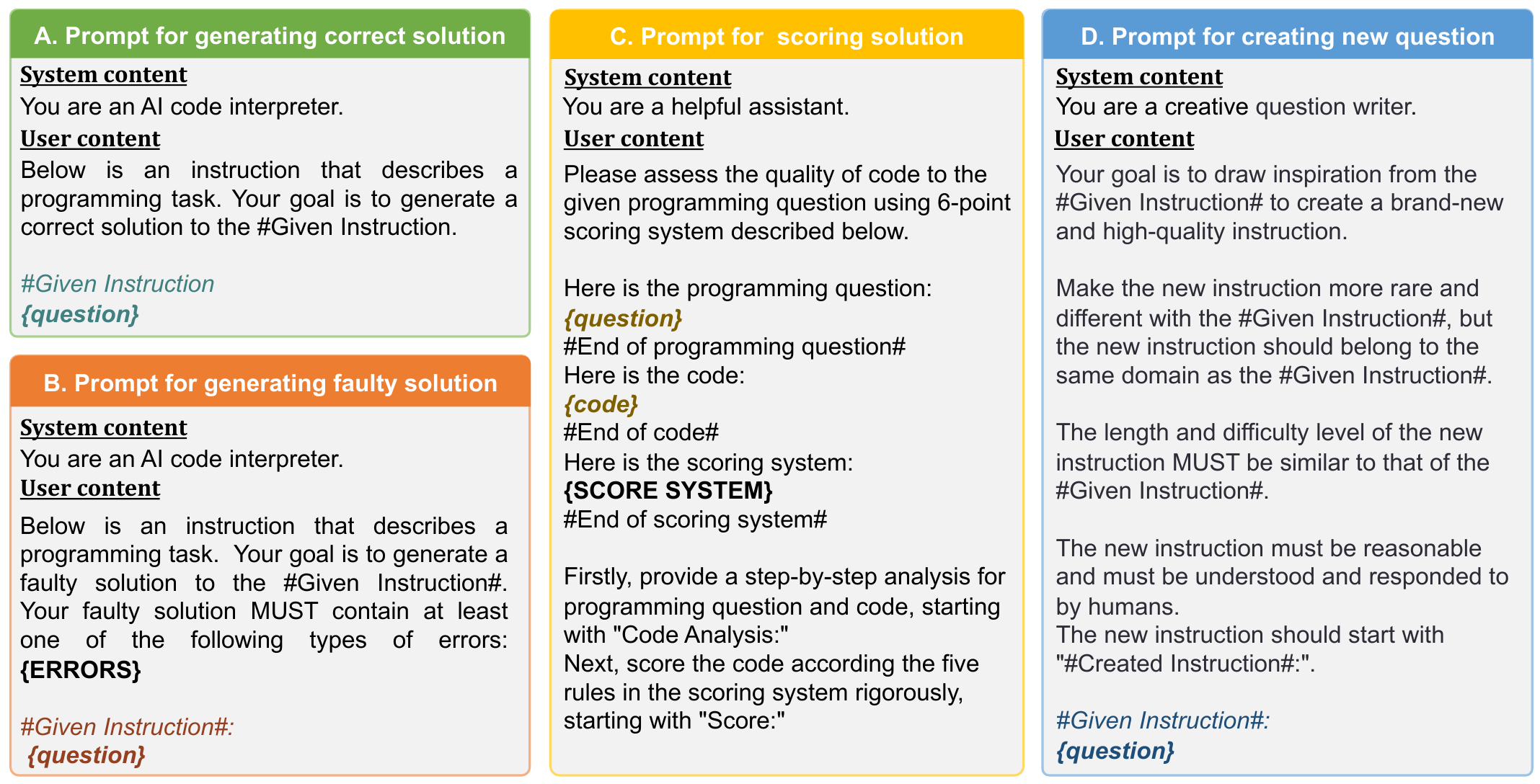}
    \caption{Prompt templates used in {\tool}.}
    \label{fig:prompt}
\end{figure}

Specifically, \add{for correct solutions, we develop the prompt template $P_c(\cdot)$ building upon the fine-tuning prompt from prior work~\cite{WizardCoder}. We enhance this base prompt by adding system content and modifying instructions to better elicit the teacher model's advanced programming capabilities, as shown in Fig.~\ref{fig:prompt}A. To validate the effectiveness of this prompt design, we invite four developers with 5+ years of experience to evaluate 200 sampled solutions. Through code execution and manual review, all solutions are confirmed to be syntactically correct and functionally complete. For faulty solutions, we create prompt template $P_f(\cdot)$ inspired by the interpreter prompt proposed in prior work~\cite{OpenCodeInterpreter}}. Following Zheng et al.~\cite{OpenCodeInterpreter}'s work, \add{we incorporate five common error types that are widely representative of typical programming issues into this template}, with details as below:

\begin{itemize}
    \item \textbf{Syntax Error}: Violations of programming language grammar, such as unmatched parentheses or misspelled keywords.
    \item \textbf{Logical Error}: Misunderstandings of the problem that lead to incorrect outcomes even though the code runs.
    \item \textbf{Type Error}: Inappropriate operations on objects of incorrect types, such as concatenating strings directly with integers.
    \item \textbf{Name Error}: Use of variables or functions that have not been defined.
    \item \textbf{Timeout Error}: Creation of infinite loops due to  conditions that can never be fulfilled, 
    potentially making the code stuck.
\end{itemize}

\begin{figure}[t]
  \centering
       \setlength{\abovecaptionskip}{-0.01em}
        \setlength{\belowcaptionskip}{-0.1em}
  \includegraphics[scale=0.39]{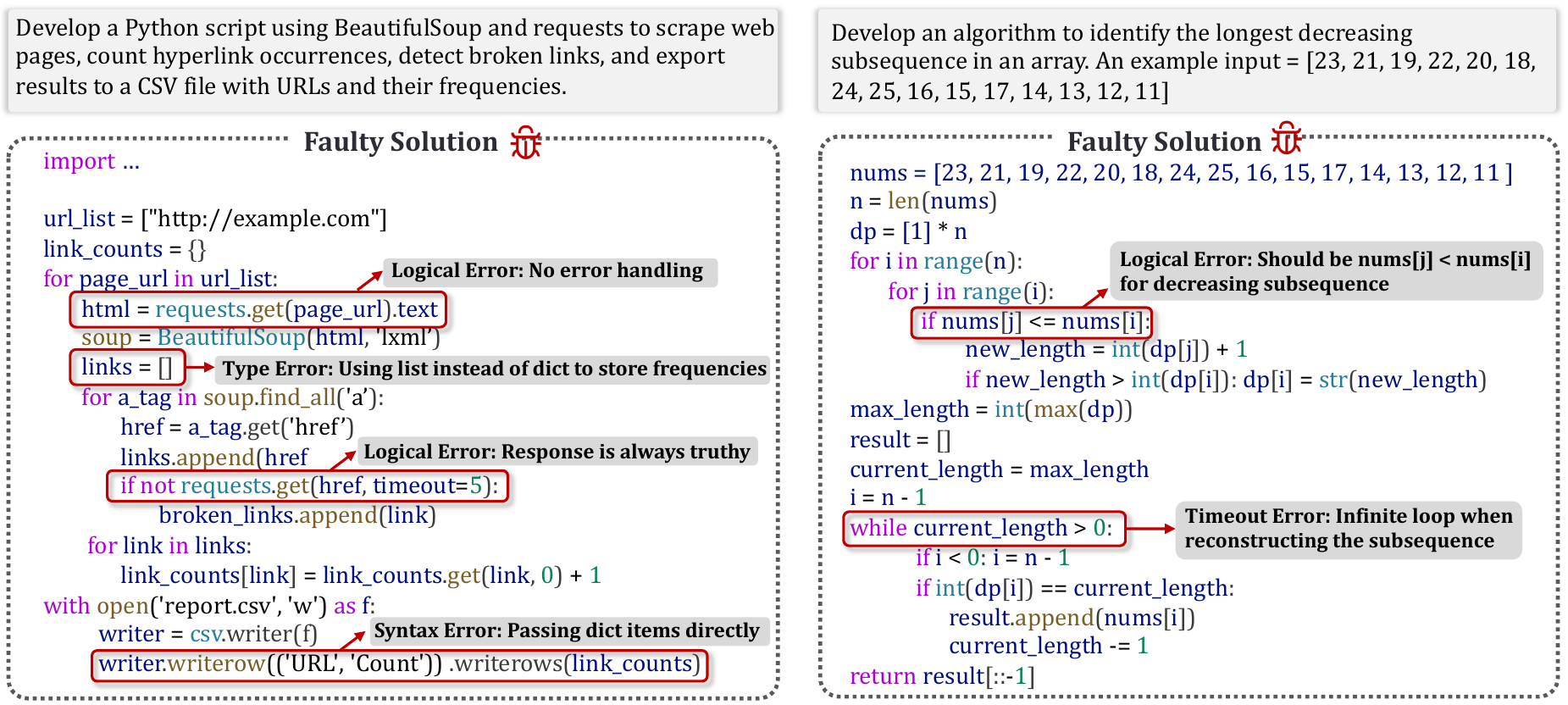}
  \caption{Two examples of faulty solutions generated by the teacher model Llama3-70B.}
  \label{fig:faulty-solution}
\end{figure}

\add{As shown in Fig.~\ref{fig:prompt}B, the prompt guides the teacher model to first understand the given programming problem, then deliberately inject specific errors based on the provided fault taxonomy. To validate the effectiveness of this prompt design, we invite four developers with 5+ years of experience to independently check 200 sampled solutions through code execution and code review. The results reveal that all solutions contain errors and cover all five error categories identified in our taxonomy. We present two examples in Fig.~\ref{fig:faulty-solution} to demonstrate various error types, including syntax errors in CSV writing, type errors in data structure usage for the web scraping task, as well as logical errors in array comparison and timeout errors due to infinite loops in a dynamic programming problem. Complete validation results are available in our GitHub repository. Using the two carefully designed prompts $P_c(x)$ and $P_f(x)$}, for each question $x$ in the seed knowledge $\mathcal{Q}=\{x^{(i)}\}_{i=1}^{n}$, the teacher generates one correct solution $y_c \sim \mathcal{T}( P_c(x) )$ and one corresponding faulty solution $y_f \sim \mathcal{T}( P_f(x) )$.

\subsubsection{Correctness-aware Supervised Learning}

This module aims to establish basic programming skills of the student model by delivering correct coding practices from the teacher model. Specifically, we fine-tune the student using the dataset $\mathcal{D_{\text{CSL}}}=\{(x^{(i)},y_c^{(i)})\}_{i=1}^{n}$, which includes programming questions paired with their correct solutions generated by the teacher. The learning objective is framed as minimizing the negative log-likelihood:
\begin{equation}
    \mathcal{L}_{CSL} = - \mathbb{E}_{(x,y)\sim\mathcal{D}_{\text{CSL}}}\sum_{t=1}^{T}\log\mathcal{S}_{\text{cls}}(y_c^{(t)}|y_c^{(<t)},x),
\end{equation}
where $\mathcal{S}_{\text{cls}}$ denotes the fine-tuned student model $\mathcal{S}(x;\theta^{\mathcal{S}}_{\text{csl}})$.
By minimizing the discrepancy between its outputs and those of the teacher model, the student model is able to 
to solve problems in a manner similar to that of
the teacher model. This stage 
lays the foundation for the subsequent 
learning phases within the framework.

\subsubsection{Fault-aware Contrastive Learning}
This module aims to further improve the student model's ability to recognize
errors.
Specifically, we fine-tune the student model $\mathcal{S}(x;\theta^{\mathcal{S}}_{\text{csl}})$ using the dataset $\mathcal{D_{\text{FCL}}}=\{(x^{(i)},y_c^{(i)},y_f^{(i)})\}_{i=1}^{m}$, in which each programming question is paired with both correct and faulty solutions generated by the teacher model. The learning objective is to achieve a higher likelihood of selecting correct solutions and a lower likelihood of selecting incorrect ones, as defined by the following:
\begin{small}
    \begin{equation}
    \mathcal{L}_{FCL} = - \mathbb{E}_{(x,y_c,y_f)\sim\mathcal{D}_{\text{FCL}}} \left[ \log\sigma \left( \beta\log\frac{\mathcal{S}_{\text{fcl}}(x,y_c)}{\mathcal{S}_{\text{cls}}(x,y_c)}-\beta\log\frac{\mathcal{S}_{\text{fcl}}(x,y_f)}{\mathcal{S}_{\text{cls}}(x,y_f)} \right) \right],
\end{equation}
\end{small}

\noindent where $\sigma$ is the logistic function, and $\beta$ is a tuning parameter that balances sensitivity to the differences between correct and faulty solutions. By distinguishing between correct and faulty solutions, the student model is able to refine 
its error identification skills.



\subsection{\secondstage} \label{sec:feedback}

As shown in 
Figure~\ref{fig:framework}, the multi-view feedback stage examines the student model's programming skills in  
addressing the original questions\modify{, for identifying the difficult questions.}
Specifically, given the questions $\{ x^{(i)}\}_{i=1}^{n}$ in seed knowledge $\mathcal{Q}$, the student model generates their solutions $\{ y_s^{(i)} \sim \mathcal{S}(P_c(x^{(i)})) \}_{i=1}^{n}$. These solutions are then evaluated by both model-based scoring and static tool-based execution, aiming at providing \modify{multi-view}
quality assessment. 


\subsubsection{Model-based Scoring}

\add{To evaluate} the solutions generated by the student model, \add{we develop a scoring prompt template based on the LLM-as-Judge prompt proposed in prior work~\cite{WizardLM}}. \add{The template incorporates five comprehensive code review standards in prior work~\cite{OpenCodeInterpreter}, with specific scoring criteria:}


\begin{itemize}[leftmargin=*]
 \item \textbf{Basic Functionality (2 Points)}: Check if the solution has obvious bugs (such as syntax and logical errors) and provides
 a basic attempt to solve the programming question.
 
 \item \textbf{Boundary Check (1 Point)}: Check if the solution includes necessary boundary checks, ensuring that the code behaves correctly under different scenarios and input ranges.
 
 \item \textbf{Requirement Coverage (1 Point)}: Check if the solution satisfies all requirements outlined in the question and successfully addresses the core requirements.
 
 \item \textbf{Readability and Documentation (1 Point)}: Check the readability of the solution, where the key part should have detailed comments for ease of understanding for other developers.
 
 \item \textbf{Coding Practices (1 Point)}: Check for efficiency, error handling, and scalability in the solution, which reflects best coding practices.
\end{itemize}
\add{To validate the effectiveness of the scoring template, we randomly sample 40 scoring results and also invite four developers with 5+ years of experience to verify whether the model: (1) strictly follows the scoring criteria in its analysis, (2) provides clear and structured evaluation processes, and (3) maintains consistency between detailed analyses and final scores. Based on human feedback, we refine the scoring prompt template step by step. After four rounds of refinement, we finally obtain a template that not only provides clear task instructions but also includes comprehensive scoring criteria and detailed analysis guidelines, as shown in Fig.~\ref{fig:prompt}C. Due to space limitations, we release the detailed results from each iteration in our GitHub repository.}

\add{Using this validated scoring prompt}, we \add{need to} annotate a small-scale seed dataset. 
\add{Given the high cost of human annotation, we use GPT-4, one of the most advanced LLM, for data annotation, as it is widely adopted for similar annotation tasks~\cite{WhatMakes}.
we invite two senior developers (with 5+ years of experience) to validate the effectiveness of this process.
Specifically, we first sample 100 question-solution pairs, then score the generated solutions by prompting GPT-4 using the scoring template and asking the expert developers to score the same samples based on the provided criteria respectively.
Based on the results, we find that GPT-4 achieves agreement with the human experts in terms of perfect accuracy in output format (100/100), and near-perfect accuracy in both code analysis (97/100) and scoring (97/100).
Thus, we choose to utilize GPT-4 for annotating the seed dataset $\mathcal{D}_\text{seed}=\{(x^{(i)},y_s^{(i)})\}_{i=1}^k$, as it maintains high-quality annotations while significantly reducing manual annotation effort.}
The labeled dataset $\mathcal{D}_\text{score}=\{(x^{(i)},y_s^{(i)},m_s^{(i)}\}_{i=1}^k$, is then used to train a CodeLlama-7B model, resulting in the scoring model. Based on the scoring model, we score all solutions generated by the student model. Each solution is scored on a scale from 1 to 6, where higher scores indicate better
code quality. \modify{In this paper, we regard solutions scored above 4 points as high-quality and 
solutions below 2 points as low-quality.}


\subsubsection{Static Tool-based Execution}

This module evaluates the executability of solutions $y_s$ within a multi-language workspace supporting Python, Java, JavaScript, C, C++, and Go; and the execution results are categorized as ``Pass'' or ``Fail''. Specifically, given a generated solution, we first extract the code snippet and then execute the code on the corresponding programming language compiler and executor. Finally, if the code can be successfully executed without errors, it is considered a ``Pass''; otherwise, it ``Fails''.
%
Note that we ignore the environment issues in the code, such as the import errors in Python (i.e., ``\textit{Failed: No module named [Libaray]}'' and name mismatching in Java (i.e., ``\textit{error: class [Name] is public, should be declared in a file named [Name].java}''). 


\subsection{\thirdstage}

As shown in Stage III of Figure~\ref{fig:framework}, the teacher model first categorizes the original questions into different difficulty levels based on the feedback in the second
stage and then generates new questions by sampling from each difficulty level
for the subsequent distillation
cycle.
%
Here, based on the \modify{model-based scores} and execution results, we categorize the questions into the following three levels:
\begin{itemize}[leftmargin=*]
    \item \textbf{Easy Question}: execution result of ``Pass'' and \modify{model-based score} $>=4$, which indicates the student can easily solve these questions with high-quality solutions.
    \item \textbf{Medium Question}: execution result of ``Pass'' and \modify{model-based score} $2\sim4$, which indicates the student can meet the basic requirements of questions, but for the code quality, there is room for improvement.
    \item \textbf{Hard Question}: execution result of ``Fail'' and \modify{model-based score} $<2$, which indicates the student struggles with these questions.
\end{itemize}

Then, to render the new questions to cover various difficulty levels, we propose \add{a template based on the in-breadth evolving prompt proposed in prior work~\cite{WizardLM}. Our template incorporates specific instructions for different difficulty levels:} 1) For the \textit{hard} and \textit{medium} levels, new questions are required to share similar topics and difficulties to the older ones. 2) For the \textit{easy} level, new questions should be more challenging than the older ones. \add{To validate the effectiveness of the template, we invite four developers with 5+ years of experience to evaluate 120 newly generated sample questions, ensuring a balanced distribution of 40 questions per difficulty category.
Specifically, we ask these developers to evaluate the questions from three perspectives: clarity of the statement, appropriateness of the difficulty level, and relevance to the original problem domain.
Then, based on human feedback, we refine the template through two iterative optimizations. The final prompt ensures that the generated questions meet the three aforementioned criteria, shown in Fig.~\ref{fig:prompt}D. Due to space limitations, we release the detailed results from each iteration in our GitHub repository.}
%
By performing the distillation process
iteratively, the student model is continuously refined through distinguishing errors and learning more advanced programming skills from the teacher model.

\section{Experimental Setup} \label{sec:setup}
    
\subsection{Research Questions} \label{sec:RQ}

In this paper, we mainly investigate the following research questions through experiments.

\begin{itemize}[leftmargin=*]

\item {\textbf{RQ1:}} How effective is {\tool} compared with the state-of-the-art knowledge distillation approaches?
\item {\textbf{RQ2:}} What is the impact 
of the four modules (i.e., \textit{faulty-aware contrastive learning}, \textit{model-based scoring}, \textit{static tool-based execution}, \textit{feedback-based classification}) on the performance of 
{\tool}?
\item {\textbf{RQ3:}} How does {\tool} perform under different parameter settings?
\item {\textbf{RQ4:}} What is the performance of {\model} compared with existing LCMs?
\end{itemize}

\subsection{Baselines} \label{sec:baseline}

To demonstrate the effectiveness of the proposed \tool,
we compare
with the following baselines.

\begin{itemize}[leftmargin=*]

\item {\textbf{EVOL-INSTRUCT~\cite{WizardCoder}}} design five evolution rules to guide the teacher model in expanding instructions: adding new constraints, replacing common requirements, complicating the original problem, providing erroneous code, and increasing complexity. This method helps elicit more complex and diverse knowledge from the teacher model.
\item {\textbf{OSS-INSTRUCT~\cite{Magicoder}}} addresses potential biases in the teacher model by referencing a wealth of open-source code, yielding more diverse and grounded data for code generation.
\item {\textbf{PERsD~\cite{personalized}}} \modify{is a personalized distillation method in which the student attempts to solve a task first, and then the teacher provides an adaptive refinement for the student to improve.}
\item {\textbf{MFT~\cite{mftcoder}}} utilizes self-instruct~\cite{self-instruct} to prompt the teacher model to generate diverse instructions for various tasks such as code completion and text-to-code generation, and then fine-tunes the student model via multi-task learning.
\end{itemize}

\subsection{Large Code Models} \label{sec:LCMs}

To conduct a comprehensive comparison, we select two proprietary LLMs, \modify{nine pre-trained} open-source LCMs, and \modify{six KD-based} open-source LCMs.    

\begin{itemize}[leftmargin=*]

\item {\textbf{GPT-4-turbo and ChatGPT (\textit{GPT-3.5-turbo})~\cite{Chatgpt}}} developed by OpenAI, has revolutionized natural language processing, enabling new possibilities for human-AI interaction. Its impressive performance in various language tasks has established it as one of the leading LLMs in the field.

\item {\textbf{CodeGen2~\cite{codegen2}}} released by Salesforce, is trained on 0.4 Trillion tokens. This model integrates various architectures and learning methodologies, outperforming in program synthesis and understanding tasks. We employ its 7B and 16B versions in our experiments.

\item {\textbf{StarCoder2~\cite{StarCoder2}}} released by the BigCode project, is trained on over 4 Trillion tokens. This training data incorporates more than 600 different programming languages as well as text extracted from GitHub pull requests and code documentation. We utilize its 7B and 15B versions in our experiments.

\item {\textbf{CodeLlama~\cite{CodeLlama}}} released by Meta AI, is further trained on Llama 2~\cite{llama2} with 0.5 Trillion code-specific tokens. Our experiments use its 7B, 13B, and 34B versions. 

\item {\textbf{DeepseekCoder~\cite{deepseek-coder}}} released by DeepSeek AI, is a series of code-specialist LLMs trained on 2 Trillion tokens across over 80 programming languages. It achieves state-of-the-art performance among open-source code LLMs. We use its 7B and 34B versions.

\item {\textbf{CodeGemma~\cite{CodeGemma}}} released by Google, is further trained on Gemma~\cite{Gemma} with an additional 0.5 Trillion tokens. This further training focuses on improving logical and mathematical reasoning, making it particularly suitable for tasks like code completion and generation. In our experiments, we utilize the 7B version of this model.

\item {\textbf{CodeQwen1.5~\cite{codeqwen}} released by Qwen, is a specialized LCM built upon the Qwen1.5~\cite{codeqwen}, trained with around 3 Trillion tokens of code-related data. It supports an extensive repertoire of 92 programming languages and exhibits exceptional capacity in long-context understanding and generation.}

\item {\textbf{WizardCoder~\cite{WizardCoder}} leverages EVOL-INSTRUCT to obtain 78,000 instructions as seed knowledge to distill the capabilities of ChatGPT into StarCoder-15B, denoted
as WizardCoder-SC in the paper.} 

\item {\textbf{Magicoder~\cite{Magicoder}} utilizes OSS-INSTRUCT to obtain 75,000 instructions as seed knowledge to distill the capabilities of ChatGPT into CodeLlama-python-7B and DeepseekCoder-6.7B, denoted
as Magicoder-CL and Magicoder-DS, respectively.}

\item {\textbf{MFTCoder~\cite{mftcoder}} distills ChatGPT's capabilities into CodeLlama-python-13B using 1,630,322 instruction samples across five code-related tasks with multi-task fine-tuning framework. In this paper, we refer to it as MFTCoder-CL.}   

\item {\textbf{WaveCoder~\cite{wavecoder}} transfers the capabilities of GPT-4, respectively into DeepSeekCoder-6.7B using 19,915 instructions across four code-related tasks as seed knowledge, denoted
as WaveCoder-DS and WaveCoderPro-DS, respectively.}

\end{itemize}

\subsection{Benchmarks} \label{sec:benchmark}

Following prior work~\cite{WizardCoder,starcoder,StarCoder2,Magicoder,CodeGemma,CodeLlama}, we evaluate all the studied LCMs on three widely-used code benchmarks: HumanEval~\cite{humaneval} and MBPP~\cite{mbpp} for Python, and MultiPL-E~\cite{MultiPL-E} for Java, JavaScript (JS), C, C++, Go and TypeScript (TS). 

\begin{itemize}

\item \textbf{HumanEval} consists of 164 manually-written Python programming problems, each accompanied by an average of 9.6 test cases. 

\item \textbf{MBPP} consists of 378 crowd-sourced Python programming problems, designed to be solvable by entry-level programmers. Each problem consists of 3 automated test cases. 
Note that, we use the sanitized version of the original MBPP (1,000 problems), which is hand-verified by the original authors.
 
\item \textbf{MultiPL-E} is a multi-language benchmark derived by translating the programming problems from HumanEval into several other programming languages. 

\end{itemize}

\subsection{Metric} We evaluate each
model’s performance on the code generation task using the commonly-adopted Pass@k metric. The Pass@k score is the percentage of solved problems out of the total problems. A problem is considered solved if any of the $k$ generated code snippets pass all test cases. Following prior studies~\cite{WizardCoder,starcoder,StarCoder2,Magicoder,CodeGemma,CodeLlama}, we use greedy decoding to generate one sample and \add{nucleus sampling with a temperature of 0.2 and top-p of 0.95 to generate more samples. The Pass@1 directly reflects the model's ability to generate correct solutions on the first attempt, which is a critical requirement for practical applications where users expect immediate correct solutions. We use the Pass@10 score to evaluate the model's performance across multiple attempts, and the first solution score to assess the quality of the initial solution.}


\subsection{Implementation Details} \label{sec:implementation}

The training and inference processes utilize the LLaMA-Factory library, a popular unified framework designed for efficient fine-tuning and fast inference across over 100+ LLMs~\cite{depoy-locally}.

\textit{1) Training Details.} 
The initial ``Seed Knowledge'' in Stage I contains 110,000 automatically generated questions from Evol-Instruct~\cite{Magicoder}. \add{To ensure the reliability of our evaluation, following the duplication practice in existing works~\cite{starcoder,Magicoder}, we use 5-grams and a Jaccard similarity threshold of 0.7 to verify that no duplication between our training data and the docstrings or solutions from HumanEval, MBPP, and MultiPL-E benchmarks.} We partition all questions in an 8:2 ratio; 80\% with correct solutions are used for correctness-aware supervised learning and 20\% with paired solutions for fault-aware contrastive learning. To save experimental cost, we randomly annotate 15K question-solution pairs from the student model’s 110K outputs \add{using GPT-4 with the solution scoring prompt} for training the scoring model. \add{The training process uses 8 NVIDIA V100 GPUs with a batch size of 256, a max sequence length of 2048, training epochs of 3, a learning rate of 4e-5 with a Cosine scheduler, and fp16 mixed precision}. We iterate
the distillation process three
times,  
with 10K newly generated questions added at each iteration, distributed in an
1:1:2 ratio for easy, medium, and hard, respectively. The impact of these parameter settings is further discussed in Section~\ref{sec:para}. The training process employs 64 NVIDIA V100 GPUs. Training hyper-parameters include a batch size of 512, a sequence length of 2048, 3 training epochs, learning rates of 2e-5 for correctness-aware supervised learning, and 5e-7 for fault-aware contrastive learning, a Cosine learning rate scheduler, and fp16 mixed precision. \add{To ensure fair comparisons, we fine-tune all baseline methods using CodeLlama-7B as the student model and maintain consistent dataset splits. All training processes are conducted using the LLaMa-Factory framework on 64 NVIDIA V100 GPUs under the same training conditions. For training hyperparameters (e.g., epoch numbers, batch size, learning rate, scheduler, model precision, weight decay, and maximum sequence length), we directly adopt the settings reported in the original papers.}


\textit{2) Inference Details.} Following previous work~\cite{code-intelligence-2,inference, Magicoder}, during inference, we use greedy decoding to generate one sample and set the max new tokens to 512. After generation, we conduct post-processing (e.g., truncating the content beyond the solution function) to ensure the generated code can be correctly evaluated following previous work~\cite{santacoder,starcoder}. 

In RQ1, RQ2 and RQ3, we use CodeLlama-7B as the student model for its widespread use and Llama3-70B~\cite{llama3} as the teacher model for its cutting-edge performance among open-source LLMs to validate {\tool}'s performance. Inspired by the prior work~\cite{Magicoder,wavecoder}, in RQ4, we further use DeepseekCoder-6.7B as the student model to develop {\model}-DS.



\section{Evaluation} \label{sec:results}
    \subsection{Effectiveness of {\tool} (RQ1)}

\begin{table}[t]
\renewcommand\arraystretch{0.85}
\setlength{\abovecaptionskip}{-0.01cm}
\small
\caption{\add{Performance comparison between SODA and baseline methods.}}
\label{tab:res-rq1}
\centering
\begin{tabular}{c|c|cc|cccccc|>{\columncolor{gray!20}}c}
\toprule
\multirow{2}{*}{Method}        & \multirow{2}{*}{Metric} & \multicolumn{1}{c|}{HumanEval} & MBPP & \multicolumn{6}{c|}{MultiPL-E} & \multirow{-1}{*}{Avg.} \\ \cline{3-10}
                               &                         & \multicolumn{2}{c|}{Python}           & Java  & JS & C & C++ & Go & TS &                \\ \midrule
\multirow{3}{*}{Base}          & Pass@1                  & \multicolumn{1}{c|}{35.4} & 46.8 & 35.9 & 32.9 & 25.1 & 29.2 & 30.4 & 32.0 & 33.4 \\
                               & Pass@10                 & \multicolumn{1}{c|}{49.5} & 60.3 & 47.7 & 46.3 & 35.7 & 41.6 & 43.3 & 46.7 & 46.3 \\
                               & Model Score             & \multicolumn{1}{c|}{2.63} & 3.64 & 2.31 & 2.15 & 1.52 & 1.94 & 2.03 & 2.35 & 2.32 \\ \midrule
\multirow{3}{*}{\shortstack{EVOL-\\INSTRUCT}} & Pass@1                  & \multicolumn{1}{c|}{48.7} & 50.0 & 55.4 & 52.4 & 41.7 & 39.6 & 43.2 & 50.0 & 47.6 \\
                               & Pass@10                 & \multicolumn{1}{c|}{60.3} & 61.9 & 65.6 & 62.7 & 52.5 & 50.5 & 54.9 & 60.2 & 58.5 \\
                               & Model Score             & \multicolumn{1}{c|}{3.62} & 3.73 & 4.12 & 3.84 & 2.92 & 2.75 & 3.13 & 3.72 & 3.47 \\ \midrule
\multirow{3}{*}{\shortstack{OSS-\\INSTRUCT}}  & Pass@1                  & \multicolumn{1}{c|}{44.5} & 52.1 & 43.9 & 42.0 & 32.5 & 34.1 & 35.9 & 45.6 & 41.3 \\
                               & Pass@10                 & \multicolumn{1}{c|}{58.1} & 66.3 & 57.3 & 55.9 & 46.3 & 48.4 & 50.0 & 60.2 & 55.3 \\
                               & Model Score             & \multicolumn{1}{c|}{3.53} & 4.14 & 3.42 & 3.23 & 2.54 & 2.73 & 2.82 & 3.74 & 3.26 \\ \midrule
\multirow{3}{*}{PERsD}         & Pass@1                  & \multicolumn{1}{c|}{56.7} & 60.5 & 53.6 & 56.0 & 39.2 & 44.5 & 41.4 & 51.8 & 50.4 \\
                               & Pass@10                 & \multicolumn{1}{c|}{68.0} & 72.5 & 65.2 & 67.3 & 51.6 & 57.1 & 54.3 & 63.7 & 62.4 \\
                               & Model Score             & \multicolumn{1}{c|}{4.13} & 4.52 & 4.24 & 3.83 & 3.14 & 3.62 & 3.25 & 4.02 & 3.84 \\ \midrule
\multirow{3}{*}{MFT}           & Pass@1                  & \multicolumn{1}{c|}{55.4} & 56.8 & 56.7 & 53.0 & 36.1 & 44.5 & 39.6 & 47.5 & 48.7 \\
                               & Pass@10                 & \multicolumn{1}{c|}{68.4} & 69.9 & 69.5 & 66.3 & 50.2 & 58.5 & 53.5 & 61.1 & 62.2 \\
                               & Model Score             & \multicolumn{1}{c|}{4.15} & 4.32 & 4.34 & 3.93 & 2.93 & 3.42 & 3.05 & 3.64 & 3.72 \\ \midrule
\multirow{3}{*}{ \textbf{SODA}}          & Pass@1                  & \multicolumn{1}{c|}{\textbf{65.2}} & \textbf{65.2} & \textbf{59.7} & \textbf{58.5} & \textbf{44.5} & \textbf{46.9} & \textbf{48.7} & \textbf{55.5} & \textbf{55.5} \\
                               & Pass@10                 & \multicolumn{1}{c|}{\textbf{75.0}} & \textbf{75.3} & \textbf{76.2} & \textbf{73.8} & \textbf{62.1} & \textbf{65.7} & \textbf{58.5} & \textbf{70.2} & \textbf{69.6} \\
                               & Model Score             & \multicolumn{1}{c|}{\textbf{4.93}} & \textbf{4.84} & \textbf{4.64} & \textbf{4.42} & \textbf{3.73} & \textbf{3.94} & \textbf{3.45} & \textbf{4.43} & \textbf{4.29} \\ \bottomrule
\end{tabular}
\end{table}

\noindent \textbf{Experimental Design.} To answer this research question, we choose  compare {\tool} with the baselines listed in Section~\ref{sec:baseline} on the three benchmarks illustrated 
in Section~\ref{sec:benchmark}. 

\textbf{Results.} Table~\ref{tab:res-rq1} shows the \add{performance} of the base model (i.e., the student model), four baseline methods and {\tool} across the three benchmarks and different programming languages. \add{We perform Wilcoxon signed-rank tests~\cite{wilcoxon} between {\tool} and each baseline method, which indicates that the improvements of {\tool} over all baselines are statistically significant (p $<$ 0.05).} Based on Table~\ref{tab:res-rq1}, we achieve the following observations.


\textbf{(1) {\tool} effectively improves the performance of the student model.}
\add{The improvements are demonstrated consistently across three evaluation metrics. In terms of Pass@1, {\tool} greatly enhances the first-attempt accuracy, increasing the score by 84.18\% on HumanEval and 39.52\% on MBPP for Python. For other programming languages, {\tool} achieves consistent Pass@1 improvements of 60.19\% $\sim$ 77.81\% on MultiPL-E. Regarding Pass@10, {\tool} shows strong potential in generating correct solutions within multiple attempts, achieving gains of 51.51\% on HumanEval (reaching 75.0) and 24.87\% on MBPP (reaching 75.3). The model score improvements are equally noteworthy, with {\tool} achieving average scores of 4.93 and 4.84 on these benchmarks, respectively, indicating enhanced solution quality even when full correctness is not achieved. This consistent improvement across all three metrics substantiates {\tool}'s effectiveness in enhancing the student model's overall programming capabilities.}


\textbf{(2) {\tool} consistently outperforms baseline methods.} \add{Compared to existing approaches, {\tool} demonstrates comprehensive advantages across all evaluation metrics. For Pass@1, it surpasses EVOL-INSTRUCT by 33.88\% and OSS-INSTRUCT by 46.51\% on HumanEval, while maintaining an average improvement of 16.61\% and 34.39\% across all benchmarks, respectively. For Pass@10, {\tool} shows improvements of 24.37\% over EVOL-INSTRUCT and 29.08\% over OSS-INSTRUCT on HumanEval, with similar advantages maintained across other tasks. The model score comparisons reveal improvements of 36.18\% over EVOL-INSTRUCT and 39.66\% over OSS-INSTRUCT, respectively. Against advanced baselines, {\tool} maintains its superiority with average improvements of 10.11\% in Pass@1, 11.53\% in Pass@10, and 11.71\% in model score compared to the strongest baseline PERsD. These comprehensive improvements validate {\tool}'s superior capability in knowledge distillation for code generation.}

\begin{tcolorbox}[breakable,width=\linewidth-2pt,boxrule=0pt,top=4pt, bottom=4pt, left=5pt,right=5pt, colback=gray!15,colframe=gray!15]
\textbf{Answer to RQ1:} {\tool} can effectively improve the capability of less powerful LCMs by transferring the knowledge of advanced LLMs,  \add{demonstrating statistically significant improvements across different benchmarks and programming languages}.

\end{tcolorbox}

\subsection{Impact of Different Modules in {\tool} (RQ2) }
\input{datas/5.3_ablation_figure}
\noindent \textbf{Experimental Design.} For this research question, we perform ablation studies by considering the following four variants of {\tool}.

\begin{itemize}
    \item {$\text{\tool}_{\text{w/o\ fault}}$:} This variant removes the faulty-aware contrastive learning method from the first stage.
    \item {$\text{\tool}_{\text{w/o\ scoring}}$:} In this variant, solution evaluation relies exclusively on execution results from the static tool. 
    \item {$\text{\tool}_{\text{w/o\ execution}}$:} In this variant, solution evaluation is based solely on the scores provided by the \modify{scoring} model.
    \item {$\text{\tool}_{\text{w/o\ classification}}$:} This variant 
    randomly samples questions to create new ones instead of sampling from different difficulty levels.
\end{itemize}

\noindent \textbf{Results.} Figure~\ref{fig:aba} shows the ablation results of {\tool} compared with the four variants in terms of Pass@1. All variants have performance degradation. Based on the Figure~\ref{fig:aba}, we achieve the following observations.

\textbf{(1) Removing faulty-aware contrastive learning causes the noticeable performance drop}. 
Specifically, the Pass@1 of {\tool} decreases by 5.67\% on HumanEval, 9.18\% on MBPP, and 2.98\% and 8.64\% across six programming languages on MultiPL-E. This decline primarily results from the student model's depressed ability to distinguish between correct and faulty coding practices.

\textbf{(2) \modify{Both model scoring and static tool execution are beneficial for increasing the performance, among which model scoring tends to be more useful.}
}
For example, on HumanEval, the Pass@1 score decreases by 2.76\% without model scoring, versus by 1.84\% without static tool execution. For
TypeScript on MultiPL-E, the removal of model scoring leads to a decrease of 4.50\% in Pass@1 score, in contrast to 1.98\% when removing tool execution. Since the \modify{scoring} model can provide a more comprehensive analysis, the static tool primarily verifies syntactical correctness. 

\textbf{(3) Random sampling for data generation leads to performance decline.}
Specifically, the average Pass@1 score decreases by 4.67\% across all the studied benchmarks.
This decline can be attributed to that random sampling cannot
provide a diverse range of question difficulties, which hampers the model's adaptive improvement in the subsequent distillation process.

\begin{tcolorbox}[breakable,width=\linewidth-2pt,boxrule=0pt,top=4pt, bottom=4pt, left=5pt,right=5pt,colback=gray!15,colframe=gray!15]
\textbf{Answer to RQ2:}  All modules in {\tool} contribute to the performance. Removing the faulty-aware contrastive learning, model scoring, static tool execution or sampling from different difficulties leads to performance decrease. \modify{Among these, faulty-aware contrastive learning is the most important module, having the greatest impact on overall performance.}
\end{tcolorbox}

\subsection{Parameter Analysis of {\tool} (RQ3)} \label{sec:para}
\definecolor{c1}{RGB}{255,196,115} 
\definecolor{c2}{RGB}{178,37,42} 
\definecolor{c3}{RGB}{103,150,118} 
\definecolor{c4}{RGB}{42,99,172}

\pgfplotstableread[row sep=\\,col sep=&]{
	k & python & java & cpp & go  \\     
	1 & 35.4 & 35.9 & 29.2 & 30.4   \\
	2 & 56.7 & 56.5 & 45.8 & 44.5 \\
	3 & 62.1 & 57.9 & 46.0 & 46.9 \\
	  4 & 64.0 & 59.7 & 46.6  & 48.7 \\
        5 & 65.2 & 59.7 & 46.9  & 48.7 \\
}\Iter

\pgfplotstableread[row sep=\\,col sep=&]{
	k & python & java & cpp & go  \\
	  6 & 59.1 & 56.0 & 40.8 & 42.6   \\
	14 & 60.3 & 57.3 & 42.0 & 43.9 \\
	22 & 65.2 & 59.7 & 46.9  & 48.7 \\
	  30 & 62.8 & 58.5 & 44.5 & 46.3 \\
        38 & 61.5 & 57.9 & 43.2 & 45.1 \\
        46 & 64.0 & 58.9 & 45.7 & 47.5 \\
}\Ratio


\begin{figure*}[t]
       \setlength{\abovecaptionskip}{-0.01em}
        \setlength{\belowcaptionskip}{-0.2em}
   \subfigure[The effect of training iterations]{
        \begin{tikzpicture}[scale=0.45]
            \huge
            \begin{axis}[
                legend style = {
                    legend columns=-1,
                    draw=none,
                },
                width=0.65\textwidth,
    		height=0.53\textwidth,
                xtick = {1,2,3,4,5},
                xticklabels = {0,1,2,3,4},
                ymin=25,ymax=73,
                ytick = {25,35,45,55,65},
                mark size=2.5pt, 
                ylabel={\bf Pass@1},
                xlabel={\bf Iteration}, 
                every axis plot/.append style={line width = 3.5pt},
                every axis/.append style={line width = 2pt},
                ]
                \addplot [mark=diamond,color=c2] table[x=k,y=python]{\Iter};
                \addplot [mark=pentagon,color=c1] table[x=k,y=java]{\Iter};
                \addplot [mark=o,color=c3] table[x=k,y=go]{\Iter};	
                \addplot [mark=square*,color=c4] table[x=k,y=cpp]{\Iter};	
                \legend{ \Large{Python},\Large{Java}, \Large{Go}, \Large{C++}}
            \end{axis}
        \end{tikzpicture}
        \label{fig:para_it}
   }
   %
   \subfigure[The effect of difficulty ratios]{
        \begin{tikzpicture}[scale=0.45]
          \begin{axis}[
            ybar=0pt,
            grid = major,
            bar width=0.56cm,
            width=1.25\textwidth,
		height=0.55\textwidth,
            xlabel={\huge \bf Ratio (easy:medium:hard)}, 
                xmin=2.0,xmax=50.0,
                xtick=data,	xticklabels= {\huge 4:0:0,\huge 2:2:0,\huge 1:1:2,\huge 1:2:1,\huge 0:4:0,\huge 0:0:4},
            legend style = {
                legend columns=-1,
                draw=none,
            },
            legend image code/.code={
                \draw [#1] (0cm,-0.18cm) rectangle (0.8cm,0.08cm); },
            ytick = {40,45,50,55,60,65},
            ymin=35,ymax=70,
            tick align=inside,
            ticklabel style={font=\huge},
          every axis plot/.append style={line width = 1.5pt},
            every axis/.append style={line width = 2pt},
            ylabel={\huge \textbf{Pass@1}},
            ]
            \addplot[pattern = north east lines,pattern color=c2] table[x=k,y=python]{\Ratio};
            \addplot[pattern =north east lines ,pattern color=c1] table[x=k,y=java]{\Ratio};
            \addplot[pattern = north east lines,pattern color=c3] table[x=k,y=go]{\Ratio};
            \addplot[pattern = north east lines,pattern color=c4] 
            table[x=k,y=cpp]{\Ratio};
  
            \legend{ \Large Python, \Large Java,  \Large Go,  \Large C++ }
            
        \end{axis}
      \end{tikzpicture}
    \label{fig:para_ratio}
   }
    \caption{Parameter analysis of {\tool}.}
    \label{fig:para}
\end{figure*}
\noindent \textbf{Experimental Design.} To answer this question, we study the impact of two parameters on the model performance,
including the iterations of distillation
and the ratio of difficulties \modify{of newly-generated seed knowledge}. 

\noindent \textbf{Results.} \modify{Due to space limitations, in Figure~\ref{fig:para}, we only show the performance of {\tool} on four popular programming languages, i.e., Python, Java, Go and C++. Based on the Figure~\ref{fig:para}, we achieve the following observations.}

\textbf{(1) {\tool} consistently improves the student model's performance as the iterations progress.} 
For example, for Python, the Pass@1 score leaps from 35.4 to 56.7, presenting a 21.3-point increase in the first iteration. It then rises modestly to 62.1 in the second iteration and continues to grow incrementally, reaching 65.2 by the fourth iteration. The results show a remarkable improvement in the first iteration and incrementally smaller gains in subsequent iterations. This may be due to the model approaching its performance saturation during the iterative distillation
process.
Consequently, considering both the performance improvement and experimental overhead, we set the number of iterations to three. 

\textbf{(2) A balanced ratio with a higher proportion for hard questions
is more
effective.} The highest performance is observed at the ratio 1:1:2 (easy:medium:hard) across all the four programming languages. 
When the ratio shifts from 1:1:2 to 4:0:0 (all easy questions), the performance decreases across all the languages. We can observe a similar trend when the ratio changes from 1:1:2 to 0:4:0 (all medium questions). Besides, when the ratio is 0:0:4 (all hard questions), the average Pass@1 score declines by 2.05\% across the four programming languages, indicating that exclusively focusing on adding hard questions also hinders continually refining the student model.
\begin{tcolorbox}[breakable,width=\linewidth-2pt,boxrule=0pt,top=4pt, bottom=4pt, left=5pt,right=5pt, colback=gray!15,colframe=gray!15]
\textbf{Answer to RQ3:} \modify{Both iterations of distillation and the ratio of difficulties of newly-generated seed knowledge have impacts on the performance of {\tool}. Specifically,} {\tool} shows substantial performance improvement initially, with more modest gains in subsequent iterations. A balanced mix of question difficulties, with a higher proportion of hard questions (1:1:2), is \modify{more} effective for improving model performance.
\end{tcolorbox}

\subsection{Performance of {\model} (RQ\add{4})}
\begin{table}[t]
\scriptsize
\setlength{\abovecaptionskip}{-0.01cm}
\caption{\add{Performance comparison between {\model} and different LLMs/LCMs. ``P1'' denotes ``Pass@1'', ``P10'' denotes ``Pass@10'', ``MS'' denotes ``Model Score''}, the abbreviations ``SC'', ``CL'' and ``DS'' refer to the student models StarCoder, CodeLlama and DeepseekCoder, respectively.}
\label{tab:res-rq2}
\centering
\begin{tabularx}{\textwidth}{c|c|XXX|XXX|XXX|XXX|XXX|XXX|XXX|XXX}
\toprule
\multicolumn{1}{c|}{Model} &{Size} & \multicolumn{3}{c|}{Python} & \multicolumn{3}{c|}{Java} & \multicolumn{3}{c|}{JS} & \multicolumn{3}{c|}{C} & \multicolumn{3}{c|}{C++} & \multicolumn{3}{c|}{Go} & \multicolumn{3}{c|}{TS} & \multicolumn{3}{c}{Avg.} \\ \midrule
 &  & P1 & P10 &MS & P1 & P10 &MS & P1 & P10 &MS & P1 & P10 &MS & P1 & P10 &MS & P1 & P10 &MS & P1 & P10 &MS & P1 & P10 & MS \\ \midrule
 \midrule
\multicolumn{26}{c}{Proprietary Models} \\ \midrule
GPT-4-Turbo & - & 90.2 & 97.7 & 6.0 & 78.5 & 86.4 & 5.4 & 75.0 & 82.5 & 5.2 & 55.2 & 63.5 & 4.2 & 62.1 & 70.8 & 4.5 & 71.9 & 79.1 & 5.1 & 74.6 & 82.1 & 5.6 & 72.5 & 80.3 & 5.1\\ 
ChatGPT& -  & 76.8 & 84.5 & 5.3 & 64.2 & 71.6 & 4.7 & 61.1 & 68.4 & 4.5 & 44.5 & 52.1 & 3.5 & 57.3 & 64.9 & 4.2 & 62.2 & 69.7 & 4.6 & 63.4 & 70.7 & 4.7 & 61.4 & 68.8 & 4.5\\ \midrule
 \midrule
\multicolumn{26}{c}{Pre-trained Open-Source Models} \\ \midrule 
Llama3 & 70B & 76.8 & 88.3 & 5.5 &  71.3 & 85.6 & 5.3 &  62.8 & 75.4 & 5.0 &  58.5 & 70.2 & 4.8 &  58.9 & 70.7 & 4.6 &  60.9 & 73.1 & 4.8 &  66.0 & 79.2 & 5.1 & 65.0 & 77.5 & 5.0   \\ \midrule
StarCoder   & 15B & 34.1 & 43.2 & 3.0 & 30.2 & 38.7 & 2.7 & 30.8 & 39.5 & 2.8 & 26.0 & 33.1 & 2.3 & 31.5 & 39.8 & 2.8 & 32.2 & 40.5 & 2.9 & 17.6 & 23.4 & 1.7 & 28.9 & 36.9 & 2.6 \\
CodeGen2  & 16B & 19.5 & 26.3 & 1.8 & 21.9 & 29.7 & 2.1 & 21.3 & 28.5 & 2.0 & 18.4 & 24.9 & 1.7 & 21.9 & 29.1 & 2.1 & 17.6 & 23.9 & 1.7 & 20.3 & 27.4 & 1.9 & 20.1 & 27.1 & 1.9  \\
CodeLlama  & 13B & 42.7 & 53.9 & 3.4 & 42.6 & 53.2 & 3.4 & 37.1 & 46.8 & 3.0 & 32.5 & 41.3 & 2.7 & 32.9 & 41.7 & 2.7 & 33.5 & 42.6 & 2.8 & 43.2 & 53.8 & 3.5 & 37.8 & 47.6 & 3.1 \\
StarCoder2 & 15B & 37.8 & 47.1 & 3.1 & 44.5 & 54.3 & 3.7 & 30.4 & 38.9 & 2.5 & 40.4 & 49.7 & 3.3 & 46.9 & 56.4 & 3.9 & 39.0 & 48.2 & 3.3 & 40.1 & 49.3 & 3.3 & 39.9 & 49.1 & 3.4    \\\midrule
CodeLlama & 7B & 35.4 & 44.7 & 2.8 & 35.9 & 44.8 & 2.9 & 32.9 & 41.2 & 2.7 & 25.1 & 32.6 & 2.0 & 29.2 & 37.5 & 2.4 & 30.4 & 38.7 & 2.5 & 32.0 & 40.1 & 2.6 & 31.6 & 39.9 & 2.6   \\
DeepseekCoder  & 6.7B & 47.6 & 54.3 & 3.5 & 43.0 & 49.2 & 3.2 & 48.4 & 54.6 & 3.5 & 42.9 & 49.1 & 3.2 & 45.1 & 51.8 & 3.5 & 44.5 & 50.3 & 3.4 & 49.7 & 56.2 & 3.7 & 45.9 & 52.2 & 3.4 \\
StarCoder2  & 7B & 35.4 & 44.2 & 3.1 & 41.4 & 50.5 & 3.3 & 31.7 & 39.8 & 2.9 & 31.9 & 40.1 & 2.3 & 34.1 & 42.6 & 3.1 & 29.2 & 37.4 & 2.7 & 35.1 & 43.7 & 3.2 & 34.1 & 42.6 & 2.9  \\
CodeGemma  &  7B&48.4	&54.1	&3.5	&48.4	&57.7	&3.2	&46.0&	51.8&	3.3	&42.2&	48.3&	3.0	&41.4&	50.5	&3.7	&34.2	&42.9	&2.4	&48.7	&54.6	&3.6&	44.1&	51.4	&3.2  \\
CodeQwen1.5  &7B  & 51.8 & 59.6 & 3.7 & 42.4 & 48.8 & 3.0 & 49.7 & 57.2 & 3.7 & 48.0 & 55.2 & 3.5 & 52.2 & 59.5 & 3.9 & 39.6 & 45.5 & 2.9 & 52.2 & 59.4 & 3.9 & 47.9 & 55.0 & 3.5 \\ \midrule
 \midrule
\multicolumn{26}{c}{KD-based Open-Source Models} \\  \midrule
WizardCoder-SC &15B & 57.3 & 61.5 & 4.1 & 35.8 & 51.5 & 3.2 & 41.9 & 50.3 & 3.4 & 37.4 & 46.2 & 3.2 & 39.0 & 48.1 & 3.3 & 37.2 & 50.0 & 3.5 & 39.5 & 48.9 & 3.4 & 41.1 & 50.9 & 3.4   \\
MTFCoder-CL  &13B & 60.3 & 66.4 & 4.4 & 57.3 & 68.2 & 4.3 & 54.2 & 63.8 & 4.1 & 44.5 & 53.4 & 3.6 & 46.3 & 55.6 & 3.9 & 45.7 & 48.1 & 3.5 & 53.0 & 62.1 & 4.4 & 51.6 & 59.7 & 4.0   \\ \midrule
Magicoder-CL  & 7B & 60.4 & 68.2 & 4.5 & 36.4 & 62.8 & 3.9 & 45.9 & 54.1 & 3.8 & 38.0 & 46.8 & 3.3 & 36.5 & 44.2 & 3.1 & 30.4 & 46.3 & 3.0 & 41.9 & 50.2 & 3.6 & 41.3 & 53.2 & 3.6    \\
Magicoder-DS & 6.7B   & 66.5 & 73.7 & 4.8 & 56.7 & 71.9 & 4.4 & 50.6 & 59.8 & 4.0 & 42.9 & 51.5 & 3.6 & 53.6 & 63.1 & 4.4 & 50.6 & 60.9 & 4.1 & 52.4 & 61.7 & 4.3 & 53.3 & 63.2 & 4.2  \\
WaveCoder-DS  & 6.7B  & 64.0 & 70.7 & 4.5 & 64.0 & 70.1 & 4.5 & 54.8 & 63.3 & 4.1 & 46.0 & 54.2 & 3.6 & 47.5 & 55.9 & 3.8 & 53.6 & 59.1 & 4.0 & 56.1 & 64.2 & 4.4 & 55.1 & 62.5 & 4.1 \\
WaveCoderPro-DS  & 6.7B  & 72.0 & 76.8 & 5.0 & 62.8 & 73.1 & 4.6 & 57.9 & 66.8 & 4.4 & 53.3 & 62.1 & 4.2 & 56.0 & 64.9 & 4.4 & 57.3 & 59.7 & 4.0 & 54.9 & 63.2 & 4.3 & 59.2 & 66.7 & 4.4 \\ \midrule
\textbf{SodaCoder-CL} & 7B & \textbf{65.2} & \textbf{75.0} & \textbf{4.9} & \textbf{59.7} & \textbf{76.2} & \textbf{4.6} & \textbf{58.5} & \textbf{73.8} & \textbf{4.4} & \textbf{44.5} & \textbf{62.1} & \textbf{3.7} & \textbf{46.9} & \textbf{65.7} & \textbf{3.9} & \textbf{48.7} & \textbf{58.5} & \textbf{3.4} & \textbf{55.5} & \textbf{70.2} & \textbf{4.4} & \textbf{54.1} & \textbf{68.7} & \textbf{4.2}\\
\textbf{SodaCoder-DS} & 6.7B  & \textbf{74.3} & \textbf{81.7} & \textbf{5.1} & \textbf{73.1} & \textbf{82.9} & \textbf{5.0} & \textbf{66.4} & \textbf{73.2} & \textbf{4.8} & \textbf{56.4} & \textbf{63.9} & \textbf{4.3} & \textbf{62.1} & \textbf{69.5} & \textbf{4.6} & \textbf{61.5} & \textbf{69.5} & \textbf{4.4} & \textbf{71.6} & \textbf{77.4} & \textbf{5.2} & \textbf{66.5} & \textbf{74.0} & \textbf{4.8}  \\ \bottomrule
\end{tabularx}
\end{table}

\noindent \textbf{Experimental Design.} To answer this research question, we compare {\model} with LCMs listed in Section~\ref{sec:LCMs}. For open-source LCMs, we download
them from HuggingFace Hub~\cite{HuggingFace} and deploy them locally. For GPT-4-turbo and ChatGPT, we use the public APIs provided by OpenAI.

\textbf{Results.} Table~\ref{tab:res-rq2} presents the performance comparison between {\model} and various LCMs in terms of Pass@1, \add{Pass@10 and model score} on different programming languages. Based on the results in Table~\ref{tab:res-rq2}, we have the following observations.


\textbf{(1) {\model} achieves the best results among 15 LCMs under 16B.} Specifically, {\model}-CL shows remarkable \add{Pass@1} improvements over 9 pre-trained models, \add{ranging from 14.55\% to 38.49\% across different programming languages}. For the 6 KD-based LCMs, {\model}-DS achieves the highest average Pass@1 score of 66.4, outperforming other distilled models with an improvement of up to 230.3\%. \add{Similar trends are observed in Pass@10, where {\model}-DS demonstrates consistent superiority with improvements of 35.2\% $\sim$ 42.8\%, and in terms of model score, showing enhancements of 31.5\% $\sim$ 39.7\% across all programming tasks}. While {\model}-DS only has a 6.7B size, it outperforms ChatGPT and reaches 90\% of GPT-4 Turbo's performance.


\textbf{(2) Knowledge distillation is very useful to enhance LCM performance.} For instance, {\model}-DS achieves 44.89\% and 38.0\% improvement in average Pass@1 and \add{Pass@10} scores compared to its base model DeepseekCoder-6.7B, respectively. Similarly, other KD methods also enhance the performance of their corresponding base models. For instance, \add{in terms of averaged model score, Magicoder-DS increases its base model’s performance from 3.4 to 4.2, MFTCoder-CL and WizardCoder-SC achieve a 29.03\% and a 30.76\% increase over their base models.}


\textbf{(3) Different student models have varying impacts on the knowledge distillation process.} For example, {\model}-DS, utilizing DeepseekCoder-6.7B as its student model, achieves an average Pass@1 score of 66.4, substantially higher than the 54.1 score of {\model}-CL, which employs CodeLlama-7B as its student model. In addition, Magicoder-DS also outperforms Magicoder-CL by 12 points (53.3 vs. 41.3), \add{and shows similar advantages in Pass@10 (63.2 vs. 53.2) and model score (4.2 vs. 3.6)}. Besides, we observe that the top three models by average Pass@1 score all utilize DeepseekCoder-6.7B, i.e., {\model}-DS, WaveCoderPro-DS and WaveCoder-DS. These results imply that the capability of the student model indeed plays a crucial role in the distillation process. \add{This performance difference can be explained by two key aspects: (1) model architecture and (2) pre-training dataset. 
For the first aspect, both models share the same fundamental designs (Transformer~\cite{transformer} with Rotary Position Embeddings (RoPE)~\cite{Rope}, pre-normalization with RMSNorm~\cite{RMSNorm}, and SwiGLU activation~\cite{SwiGLU}), however, DeepseekCoder doubles the intermediate layer size (11008 vs 5504), attention heads (32 vs 16), and adopts FlashAttention v2~\cite{FlashAttention} for efficient attention computation. These architectural improvements appear to provide better feature extraction and representation capabilities during the knowledge distillation process. For the second aspect, DeepseekCoder is trained from scratch on 2 trillion tokens with a code-focused dataset (87\% source code, 10\% code-related English corpus, 3\% Chinese corpus), while CodeLlama is initialized from Llama 2 (trained on 2 trillion tokens of natural language) and further trained on 500B tokens of code-heavy data (85\% source code, 8\% code-related natural language, 7\% Chinese natural language). The larger-scale training and more code-centric data composition of DeepseekCoder appear to provide a stronger foundation for code understanding, leading to superior performance when used as a student model in knowledge distillation. These findings highlight the crucial role of both architecture design and pre-training datasets in the effectiveness of our method.}

\begin{tcolorbox}[breakable,width=\linewidth-2pt,boxrule=0pt,top=4pt, bottom=4pt, left=5pt,right=5pt,colback=gray!15,colframe=gray!15]
\textbf{Answer to RQ4:} {\model} achieves the best results among the 15 LCMs under 16B, outperforming the 9 pre-trained LCMs by up to 81.08\% and 6 KD-based LCMs by up to 32.19\% in terms of average Pass@1. Notably, {\model}-DS-6.7B outperforms ChatGPT and reaches 90\% of the average performance of GPT-4-turbo.
\end{tcolorbox}


\section{Discussion} \label{sec:discussion}
    
\subsection{Case Study}

\begin{figure*}[t]
 \setlength{\abovecaptionskip}{-0.01em}
        \setlength{\belowcaptionskip}{-0.2em}
    \centering
    \subfigure[HumanEval/14]{
\includegraphics[scale=0.4]{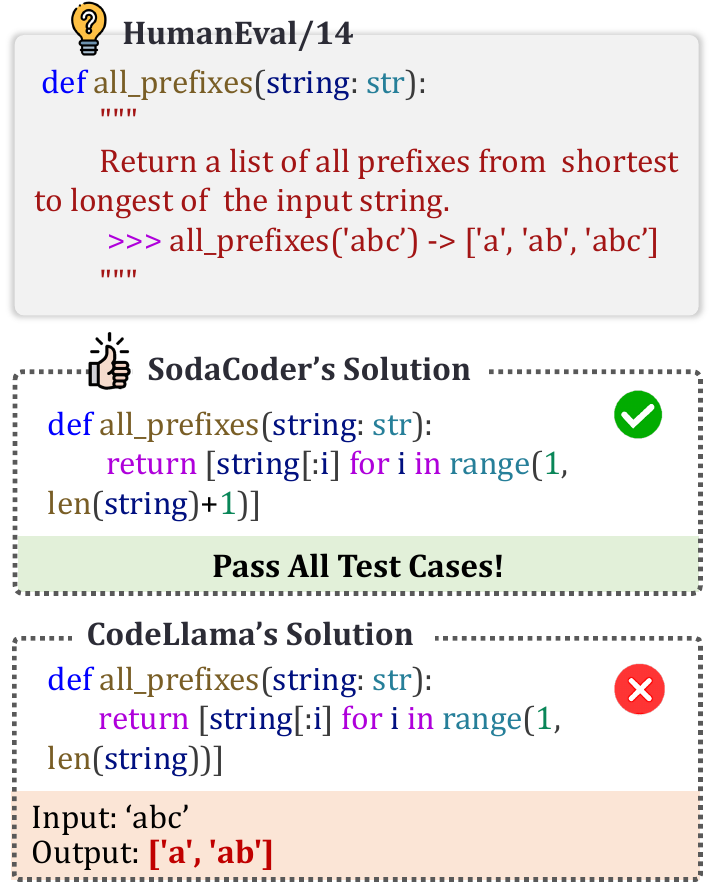}
      \label{fig:case_study_main_1}
   }
   \hspace{0.1cm}
    \subfigure[HumanEval/95]{
\includegraphics[scale=0.4]{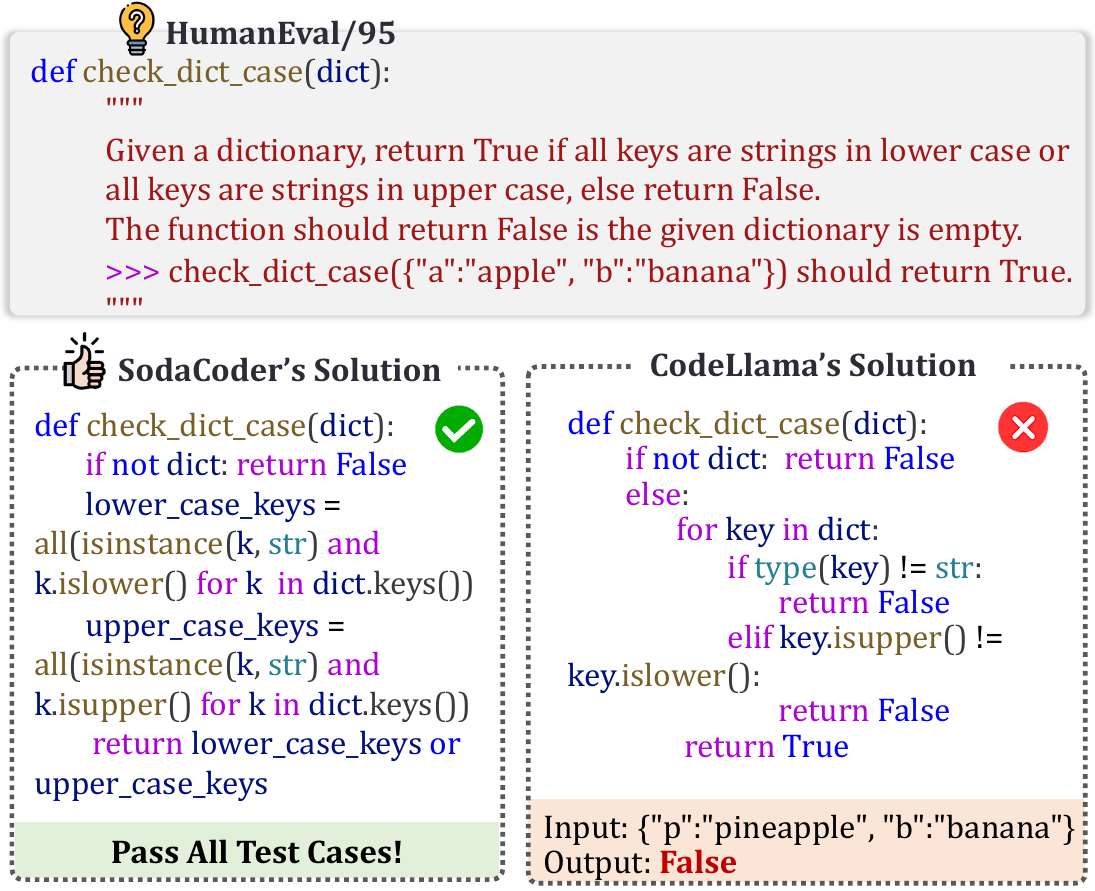}
      \label{fig:case_study_main_2}
   }
    \caption{Case studies of comparing {\model} to the corresponding student model CodeLlama.}
    \label{fig:cases}
\end{figure*}

\input{datas/6.2_analyst}

In this section, we present two cases from the HumanEval benchmark to demonstrate the enhanced error recognition and programming logic capabilities of {\model} compared to the corresponding student model, CodeLlama, as shown in Figure~\ref{fig:cases}.

\textit{Enhancement of error recognition ability.} As shown in Figure~\ref{fig:case_study_main_1}, the task is ``\textit{Return a list of all prefixes from shortest to longest of the input string}''. {\model} correctly uses the API ``\texttt{range(1, len(string)+1)}'', ensuring that the full string is included in the list of prefixes. This solution passes all test cases, indicating a solid understanding of Python's range function and string slicing. Conversely, CodeLlama incorrectly uses the API ``\texttt{range(1, len(string))}'', resulting in the omission of the full string in the list of prefixes. The output \textit{[`a', `ab']} for input \textit{`abc'} fails to pass the test case, highlighting a misunderstanding of the range function's upper boundary. The solutions demonstrate {\model}'s enhanced
error recognition ability because it has been ``teached''
to distinguish between correct and faulty coding practices.

\textit{Improvement of programming logic ability.} As shown in Figure~\ref{fig:case_study_main_2}, the task is ``\textit{Given a dictionary, return True if all keys are strings in lower case or all keys are strings in upper case, else return False}''. {\model} correctly handles mixed data types of the dictionary keys with the API ``\texttt{isinstance()}'' and employs Python's list comprehensions ``\texttt{all()}'' to effectively check case conditions. CodeLlama fails in its logical implementations, which incorrectly checks if the key's ``\texttt{isupper()}'' and ``\texttt{islower()}'' produce different results. The solutions illustrate {\model}'s improved programming logic by properly implementing the checks for all keys being either in lower or upper case, compared to CodeLlama's flawed logic.



\subsection{Effectiveness of the Scoring Model}

\subsubsection{User Study}

To evaluate the effectiveness of our scoring model, we conduct a user study with solutions of varying execution statuses.

\textit{Study Design.} We randomly sample 33 solutions from three categories: uncompilable, non-executable, and executable, totaling 99 samples. For the study, we design a questionnaire, in which each item contains a programming question, a solution, and two analyses provided by the scoring model or GPT-4. We invite 9 developers from Huawei, each with over six years of programming experience, to evaluate 11 cases in one questionnaire. They are asked to judge whether the scoring or GPT-4 provides better analysis, or if the analyses are comparable (i.e., ``\textit{Scoring Model win}'', ``\textit{GPT-4 win}'' or ``\textit{Tie}''). To ensure an unbiased evaluation, participants are not informed about which analysis is provided by which model.

\textit{Result Analysis}. Fig.~\ref{fig:user_study_1} depicts the results of our user study. Specifically, the scoring model outperforms GPT-4 in uncompilable and executable categories with 14 and 13 wins respectively, compared to GPT-4’s 13 and 10 wins. In the non-executable category, GPT-4 leads with 18 wins to the scoring model's 8, indicating GPT-4's better understanding of solutions' complex issues. Overall, the scoring model surpasses GPT-4 in 35 cases, and ties in 25, covering over 60\% of the 99 total cases. The results confirm the effectiveness of the scoring model, which can provide usable and reliable quality analysis while reducing experimental costs compared to using GPT-4.

\add{\textit{Expert Evaluation.} To further validate the reliability of the scoring rubric, we conduct an inter-rater agreement study with two senior developers (5+ years of experience). The experts independently score the same set of 99 samples following our scoring rubric, yielding a Cohen's Kappa coefficient~\cite{agreement} of 0.953 between their results, indicating excellent inter-rater reliability. Next, comparing each expert's scores with the scoring model's predictions, we obtain Cohen's Kappa values of 0.868 and 0.863, respectively, demonstrating strong agreement between our automated scoring and expert judgment. These results, along with our ablation study findings, confirm the effectiveness and reliability of our scoring rubric.}

\subsubsection{Case Study} 
Figure~\ref{fig:user_study_2} illustrates a case study of multi-view feedback from the initial iteration cycle. The student model provides a solution, attempting to solve the question about manipulating elements of an existing array to construct a new one. Although the static tool verifies that the solution is syntactically correct, the scoring model identifies several flaws. It pinpoints incorrect implementations of the prefix sum and array value calculations, which fails to address the question. Besides, the solution lacks necessary boundary checks and error handling, and it does not fulfill the specified task requirements. This case study underscores the scoring model's effectiveness in detecting logical errors and poor coding practices, thereby providing more comprehensive feedback on the student model's generation.

\subsection{Impact Analysis of Teacher LLMs}

In this experiment, we evaluate the impact of different teacher models in our {\tool} framework.
Specifically, we use the state-of-the-art LLM, GPT-4 Turbo to replace Llama3-70B as the teacher model to train the same student model CodeLlama-7B. Table~\ref{table:teacher} presents the results under these two teacher models.
We can see that Llama3-70B often matches or exceeds the performance of GPT-4 Turbo with respect to various programming languages. 
For example, Llama3-70B achieves a higher Pass@1 score than GPT-4 Turbo in terms of Python (65.2 vs. 64.0), C (44.5 vs. 39.8), and TypeScript (55.5 vs. 53.7).
As for JavaScript  (60.3 vs. 58.5), C++ (48.1 vs. 46.9), and Go (49.3 vs. 48.7), GPT-4 Turbo shows slightly better results than Llama3-70B.
More importantly,
Llama3-70B achieves a higher average
Pass@1 score (54.1) compared to GPT-4 Turbo (53.4) across all the languages. 

\begin{table}[t]
\small
\renewcommand*{\arraystretch}{0.75}
\setlength{\abovecaptionskip}{-0.01cm}
\caption{Performance comparison of different teacher LLMs in terms of Pass@1.}
\label{table:teacher}
\begin{tabular}{c|c|c|c|c|c|c|c|c|c}
\toprule
Teacher   & Type &  Python   & Java & JS  & C & C++ & Go  & TS & Avg. \\ \midrule
GPT-4 Turbo    & Proprietary  & 64.0    & 59.1   & \textbf{60.3} & 39.8 & \textbf{48.1}    & \textbf{49.3} & 53.7 & 53.4 \\
Llama3-70B     & Open-source  & \textbf{65.2}    & \textbf{59.7}  & 58.5 & \textbf{44.5} & 46.9 & 48.7 & \textbf{55.5}  & \textbf{54.1} \\ \bottomrule
\end{tabular}

\end{table}

\add{\subsection{Scaling Analysis of Fault Taxonomy}}

\begin{table}[t]
\renewcommand*{\arraystretch}{0.7}
\setlength{\abovecaptionskip}{-0.01cm}
\caption{\add{The performance comparison between KD-based models and SodaCoder on ComplexCodeEval benchmark for the test case generation task.}}
\label{tab:res-complex-codegen}
\small
\centering
\begin{tabular}{c|cc|cc}
\toprule
\multirow{2}{*}{Models} & \multicolumn{2}{c|}{Python} & \multicolumn{2}{c}{Java}   \\ \cline{2-5} 
                        & Edit Similarity  & CodeBLEU & Edit Similarity & CodeBLEU \\ \midrule
WizardCoder             & 19.74            & 10.97    & 21.53           & 14.32    \\ \midrule
MFT-Coder               & 22.87            & 12.18    & 20.80           & 13.38    \\ \midrule
Magicoder-CL            & 19.40            & 10.13    & 18.27           & 11.32    \\ \midrule
Magicoder-DS            & 22.89            & 12.40    & 22.67           & 15.33    \\ \midrule
WaveCoder-DS            & 21.36            & 11.32    & 22.21           & 13.79    \\ \midrule
WaveCoderPro-DS         & 23.48            & 12.68    & 23.66           & 15.70    \\ \midrule
SodaCoder-CL            & 26.52            & 14.43    & 25.69           & 17.17    \\
- w/o fault             & 23.21                 & 11.12         & 22.38                 & 13.86         \\ \midrule
SodaCoder-DS            & \textbf{26.54}            & \textbf{14.67}    & \textbf{26.23}           & \textbf{17.38}    \\
- w/o fault             &  24.63                & 12.36         & 23.92                & 15.49         \\ \bottomrule
\end{tabular}
\end{table}

\add{While there are only simple error patterns included in our fault-aware contrastive learning module, it still enables student models to solve complex programming tasks. This is mainly because these simple error patterns reflect common programming errors frequently encountered in real-world development. 
To validate this, we conduct experiments on the ComplexCodeEval benchmark~\cite{ComplexCodeEval} which contains more complex programming tasks in the real world. The results are shown in Table~\ref{tab:res-complex-codegen}. As we shall see, learning these basic error patterns also improves performance on complex programming tasks, with our model outperforming all baselines and removing the fault-aware contrastive learning module causing a 10.34\% average decrease in edit similarity. 
Besides, performance can be further improved by expanding our fault taxonomy to include more sophisticated error types, such as concurrency bugs, race conditions, and architectural design flaws. While fault taxonomy is valuable for enhancing the capabilities of student models, generating and managing these faults poses key challenges for large programming tasks.
For generation, as program size grows, synthesizing meaningful faults requires handling intricate dependencies across multiple components. For management, validating a diverse set of faulty solutions becomes increasingly complex, especially when dealing with distributed systems. Therefore, in the future, there is a need to develop modular fault generation techniques to handle component-level dependencies and automated validation approaches to streamline the management process.}

\add{\subsection{Cost Analysis of {\tool}}}

\subsubsection{\add{Training and Inference Costs}} \add{Using Llama3-70B as the teacher and DeepseekCoder-6.7B as the student, with 110k initial seed knowledge items, the complete training process takes approximately 58.5 hours on 64 V100 GPUs over three iterations. Each iteration comprises three stages: Stage I (12.5 hours) for solution generation and learning, Stage II (6 hours) for student examination and feedback, and Stage III (1 hour) for new question generation. While this represents great upfront computation, the long-term inference benefits are substantial. Specifically, SodaCoder-DS achieves 2.15 tokens/s throughput on 8 V100 GPUs, far exceeding Llama3-70B's 0.08 tokens/s. Our analysis shows that the breakeven point occurs after processing 435,942 tokens, equivalent to approximately 218 inferences at 2,000 tokens per inference~\cite{LongWriter}. Beyond this point, SODA delivers substantial computational savings while maintaining competitive performance, making it an especially attractive solution for deployment scenarios requiring frequent inference operations.}

\subsubsection{\add{Training Cost Comparison.}} \add{We focus on comparing the fine-tuning time of the student model CodeLlama-7B across different approaches. The results show that SODA requires 8.5 hours for fine-tuning, which is equivalent to EVOL-INSTRUCT. This duration is moderately longer than OSS-INSTRUCT and PERsD, which take 6.5 and 7 hours, respectively, and exceeds MFT's 3-hour training time. Despite the slightly increased training duration, SODA demonstrates superior performance, indicating an effective trade-off between computational cost and model capability.}

\subsection{Threats to Validity}

We identify three main threats to the validity of our study. 

\begin{itemize}[leftmargin=*]
    \item \textbf{The selection of models}. In our experiments, we select two teacher LLMs and two student LCMs. However, many other models could serve as teachers and students, such as Gemini~\cite{Gemini} and Mixtral~\cite{Mixtral} as teacher LLMs, with StarCoder and CodeGen as student LCMs. In the future, we plan to explore more combinations of teacher and student models to further assess the generalizability of our distillation framework. 

    \item \textbf{The selection of benchmarks.} We adopt widely-used code benchmarks that evaluate LLMs on generating single-function programs from natural language descriptions. While {\tool} demonstrates superior performance on these benchmarks, other benchmarks for data science (DS-1000~\cite{DS-1000}) and addressing open-source issues (SWE-bench~\cite{swe}) are not included in our experiments. In the future, we will consider these additional benchmarks to provide a more comprehensive evaluation.

     \item  \add{\textbf{The design of prompt templates.} The prompt templates are manually constructed through iterative refinement with expert validation. While the experiments demonstrate their effectiveness in prompting LLMs' output, these templates may be suboptimal for corresponding tasks. In the future, we will incorporate prompt optimization tools (e.g., DsPy~\cite{DsPy}) to automatically generate the optimal templates for different tasks.}
    
\end{itemize}

\subsection{\add{Limitations and Future Work}}

\add{While {\tool} demonstrates strong performance in knowledge distillation, we identify two main limitations: (1) Common error types in fault knowledge learning. Our fault taxonomy only includes five common error types but lacks coverage for complex, and scenario-specific errors (e.g., concurrency issues). In the future, we will include more types of fault taxonomy to make our model more applicable to real-world scenarios. (2) Training overhead in iterative question-answering. Our method incurs substantial computational overhead during question and answer generation, as each iteration requires prompting the LLM to generate new questions and answers. In the future, we plan to optimize this process through more efficient question selection and parallel processing, making SODA more accessible in resource-constrained environments.}


\section{Conclusion} \label{sec:conclusion}
    In this paper, we propose {\tool}, a novel self-paced knowledge distillation framework, which aims to adaptively transfer the programming capabilities of larger, advanced LLMs to smaller, less powerful LLMs. In {\tool}, we propose a novel knowledge delivery method combining correctness-aware supervised learning and fault-aware contrastive learning to improve the student model’s programming skills; and an adaptive knowledge update method based on multi-view feedback to continuously refine the student model. Based on the proposed \tool,
we develop {\model}, a series of lightweight yet effective LCMs. The evaluation on the code generation task across different benchmarks and
programming languages demonstrates the superior performance of {\model}. In the future, we will conduct evaluation
on more code intelligence tasks.

\section{Data Availability}

We release our source code and data at \url{https://github.com/yujiachen99/SodaCoder}.


\normalem
\bibliographystyle{ACM-Reference-Format}
\bibliography{acmart}

\end{document}